\documentclass[article,12pt]{article} 

\usepackage{amscd,amsmath,amssymb,amsfonts,latexsym,mathrsfs,amsthm}
\usepackage{graphicx} 
\usepackage{setspace} 
\usepackage{cancel}
\usepackage{graphics,epsfig}
\usepackage{sectsty}
\usepackage{natbib}
\usepackage[english]{babel}
\usepackage{amsmath}
\usepackage{amsfonts}
\usepackage{amssymb,amsthm}
\usepackage{bm}
\usepackage{mathrsfs}
\usepackage{subfigure}
\usepackage[usenames]{color}
\usepackage{rotating}

\usepackage{url}

\usepackage{flushend}
\usepackage{verbatim}
\usepackage{tabularx}
\usepackage{multirow} 
\usepackage{arydshln}
\usepackage{bm}
\usepackage{pifont}
\newcommand{\cmark}{\ding{51}}%
\newcommand{\xmark}{\ding{55}}

\def\e{{\bf e}}

\def\x{{\mathbf x}}

\def\y{{\bf y}}

\newcommand{\post}{\bar{\pi}}

\newtheorem{Remark}{\bf Remark}[]

\UseRawInputEncoding

\title{A note on the area under the likelihood and the fake evidence for model selection}
\author{L. Martino$^\star$ and F. Llorente$^\top$  \\
{\small$^\star$  Universita' di Catania, Catania, Italia.}\\
{\small $^\top$  Stony Brook University, NewYork, USA.} \\
}

\date{}
 
\begin{document}

\maketitle

\thispagestyle{empty}

\begin{abstract}

Improper priors are not allowed for the computation of the  Bayesian evidence $Z=p({\bf y})$ (a.k.a., marginal likelihood), since in this case $Z$  is not completely specified due to an arbitrary constant involved in the computation. However, in this work, we remark that they can be employed in a specific type of model selection problem: when we have several (possibly infinite) models belonging to the same parametric family (i.e., for tuning parameters of a parametric model). However, the quantities involved in this type of selection cannot be considered as Bayesian evidences: we suggest to use the name  ``fake evidences'' (or ``areas under the likelihood'' in the case of uniform improper priors). 
We also show that, in this model selection scenario, using a diffuse prior and increasing its scale parameter asymptotically to infinity, we cannot recover the value of the area under the likelihood,  obtained with a uniform improper prior. We first discuss it from a general point of view. Then we provide, as an applicative example, all the details for Bayesian regression models with nonlinear bases, considering two cases: the use of a uniform improper prior and  the use of a Gaussian prior, respectively. A numerical experiment is also provided confirming and checking all the previous statements. 
\end{abstract}

\section{Introduction}
\label{sec-intro}

Nowadays,  Bayesian inference is a hot topic of research and, as a consequence, Bayesian methods are considered more and more as benchmark techniques for inferring the parameters of a model (and their uncertainties), and/or for model selection purposes.  Although Bayesian inference has historically been always used (e.g. \citep{Liu04b,Robert04}),  Bayesian analyses are now becoming more widespread:  we can find Bayesian studies in very different applied fields such as remote sensing \citep{MartinoCompress2,DeepLLORENTE}, astronomy \citep{Anfinogentov2021,Feroz2019}, cosmology \citep{Ashton2021,Ayuso2021}, or optical spectroscopy \citep{Emmert2019,RevModPhys.83.943}, to name a few. 
\newline
In Bayesian inference, we can distinguish (at least) two levels: the inference over the parameters   (Level-1) and the model selection problem (Level 2).
In order to perform Bayesian model selection (Level-2), we need to compute the so-called {\it Bayesian evidence}, a.k.a., {\it marginal likelihood} of the model, denoted in this work as $Z$. 
The choice of the prior densities over the parameters (in Level-1) affects the value of the marginal likelihood. 
\newline
\newline
Vague/diffuse priors and, more extremely, improper priors are generally employed (when possible) in level-1 of inference for expressing a weak a-priori information (for this reason, they are also called non-informative priors)  \citep{Gregory2011,Pascoe2020}. However, in model selection (level-2), the use of vague/diffuse priors over the parameters (in level-1) can radically change the value of the evidence $Z$. Therefore, in this sense,  vague/diffuse priors are always informative in level-2. Moreover, the use of  improper priors is forbidden for computing the evidence $Z$ since, in this case, the marginal likelihood $Z$  is not completely specified due to an arbitrary constant involved in the computation.
\newline
\newline
In this work, we firstly try to clarify and remark the issues described above in order to avoid any sort of confusion in the literature \cite{Fitzgerald96_bo,LlorenteSafePriors}. Moreover, we show that although improper priors are not allowed for the computation of the evidence $Z$, they can be employed in a specific type of model selection problem: when we have several (possibly infinite) models belonging to the same parametric family (i.e., for tuning parameters of a parametric model). However,
in this case, we are {\it not} actually computing an evidence $Z$  (that is not completely specified) \cite{Fitzgerald96_bo}. For this reason, we suggest  to call the calculated quantity as ``fake evidence'' or, in the case of uniform improper priors, as ``the area under the likelihood''.  Furthermore, in this scenario, if we apply a vague/diffuse prior and leave the scale parameter to increase tending to infinity,  it is not possible to recover the results obtained by employing improper uniform prior   (as typically happens in level-1).
\newline
We show and discuss these points firstly with generic arguments, and then more specifically within a Bayesian regression model. We provide theoretical details comparing the scenario with a uniform improper prior with the case of a Gaussian prior, checking and confirming the general statements previously discussed. Moreover, a specific example of generalized linear model is considered for providing numerical checks and related simulations. 
\newline
\newline
The rest of the work is structured as follows. The main background and notation, as well as the different levels of inference  and types of model selection are given in Section \ref{ModelFitSect}. A detailed discussion about the safe use of vague and/or improper priors is given in Section \ref{FernandoMeMata}. The key observations of the work are described in Section \ref{KEYPOINT}. Section \ref{Chap6FirstSect} provides a detailed description of a regression problem with Bayesian generalized linear models considering a uniform improper prior and a Gaussian prior. Section \ref{NumEspALL} provides related numerical results. Finally, several conclusions are given in Section \ref{SectConcl}.

\section{Elements in Bayesian inference}\label{ModelFitSect}


In many applications, the goal is to make inference 
 about a variable of interest,  
${\bm \theta}=\theta_{1:D_{{\bm \theta}}}=[\theta_1,\theta_2,\ldots,\theta_{D_{{\bm \theta}}}]\in {\bm \Theta}\subseteq\mathbb{R}^{D_{{\bm \theta}}}$,
where $\theta_d\in \mathbb{R}$ for all $d=1,\ldots,D_{{\bm \theta}}$, given a set of observed measurements ${\bf y}=[y_1,\ldots,y_{D_y}]\in \mathbb{R}^{D_y}$.
The observed vector ${\bf y}$ is linked with 
the vector of parameters of interest ${\bm \theta}$ by an observation model denoted as $\mathcal{M}$, which induces a likelihood function denoted as $\ell({\bf y}|{\bm \theta},\mathcal{M})$ (that is a density with respect to ${\bf y}$ and a non-negative function fixing {\bf y} and varying ${\bm \theta}$).
\newline
In the Bayesian framework, a complete model $\mathcal{M}$ is formed by a likelihood function $\ell({\bf y}|{\bm \theta},\mathcal{M})$ and a prior probability density function (pdf) $g({\bm \theta}|\mathcal{M})$ chosen by the practitioner.  Then, all the statistical information is summarized by the posterior pdf, i.e.,
\begin{equation*}
{\bar \pi}({\bm \theta}|\y,\mathcal{M})= \frac{\ell(\y|{\bm \theta},\mathcal{M}) g({\bm \theta}|\mathcal{M})}{p(\y|\mathcal{M})},
\label{eq:posterior}
\end{equation*}
where 
\begin{equation}\label{MarginalLikelihood}
Z= p({\bf y}|\mathcal{M})=\int_{\bm \Theta} \ell({\bf y}|{\bm \theta},\mathcal{M}) g({\bm \theta}|\mathcal{M}) d{\bm \theta},
\end{equation}
is called  {\it Bayesian evidence} or {\it marginal likelihood}  \citep{Robert04,Liu04b,LlorenteREV19}. 
This quantity is important for model selection purposes, as we show below.
Usually $Z= p({\bf y}|\mathcal{M})$ is unknown and difficult to approximate, so that in many situations we are only able to evaluate the unnormalized target function,
$\pi({\bm \theta}|\y,\mathcal{M})=\ell({\bf y}|{\bm \theta},\mathcal{M}) g({\bm \theta}|\mathcal{M}) \propto {\bar \pi}({\bm \theta}|\y,\mathcal{M})$,
so that $\bar{\pi}({\bm \theta}|\y,\mathcal{M})=\frac{1}{Z}\pi({\bm \theta}|\y,\mathcal{M})$ and $Z= \int_{\bm \Theta} \pi({\bm \theta}|\y,\mathcal{M}) d{\bm \theta}$.  

\subsection{Levels in Bayesian inference}

%
\noindent
Generally speaking, in Bayesian inference we can distinguish between two types of problems or levels of inference \citep[Ch. 28]{mackay2003information}, described below:
\newline
\begin{itemize}
\item {\bf Level-1: Estimation and prediction problems.} 
In the first level, given the $m$-th model $\mathcal{M}_m$, we are interested in making inferences regarding parameter ${\bm \theta}_m$ by focusing on its posterior pdf $\post({\bm \theta}_m|\y,\mathcal{M}_m) \propto \ell(\y|{\bm \theta}_m,\mathcal{M}_m)g({\bm \theta}_m|\mathcal{M}_m)$. This is also denoted  as ``Level-1 of inference'' in the literature. Now  we drop for simplicity the dependence on the $m$-th model $\mathcal{M}_m$ and $m$, then  $\post({\bm \theta}_m|\y,\mathcal{M}_m)=\post({\bm \theta}|\y)$.

\item {\bf Level-2: Model selection problems.} 
In the second type of problem, we focus on the model posterior distribution 
$$
p(\mathcal{M}_m|\y) \propto p(\mathcal{M}_m)Z_m = p(\mathcal{M}_m)\int_{{\bm \Theta}_m}\ell(\y|{\bm \theta}_m,\mathcal{M}_m)g({\bm \theta}_m|\mathcal{M}_m)
$$
for all $m=1,\dots,M$.
 This is also known  as ``Level-2 of inference''. 
\end{itemize}
More {\it levels} of inference can be recognized in the so-called hierarchical  Bayesian approaches.  However, conceptually these are the two {\it main} levels of inference since they are associated with the two main inference scenarios: parameter estimation and model selection. We will see that the prior choice has a different impact in each of the different levels.  In this work, we focus mainly on level-2.

\subsection{Type of model comparison} \label{TypeModSection}

\noindent
In the literature, we can distinguish different types of model selection,  as we summarize below. The type of model selection problem can affect the user's choice of a suitable prior density. 
\newline
\begin{itemize}
\item  {\bf Type-1 --- Basic model selection: } In this scenario, we compare different likelihood functions (i.e., observation models). The likelihood functions can represent completely different models, living even in different parameter spaces. In this scenario, the parameters ${\bm \theta}_m$ of each model can have a completely different physical or statistical interpretation. 
\item {\bf Type-2 --- Models in the same parametric families:} tuning the parameters of a parametric model can be considered a model selection problems where different models of the same parametric families are compared. Indeed, in this case, we can apply the so-called {\it empirical Bayesian} approach.
Let consider now that the observation model  depends on some vectors of parameters ${\bm \eta}$, i.e. $\ell(\y|{\bm \theta},{\bm \eta})$.
The marginal likelihood would depend on ${\bm \eta}$,
\begin{equation}\label{MarginalLikelihood3}
Z=p(\y|{\bm \eta}) = \int_{\bm \Theta} \ell(\y|{\bm \theta},{\bm \eta}) g({\bm \theta}) d{\bm \theta}.
\end{equation}
The empirical Bayesian approach consists on tuning ${\bm \eta}$ by maximizing $p(\y|{\bm \eta}) $ keeping fixed $\y$, i.e.,
\begin{align}
{\bm \eta}^*=\arg\max p(\y|{\bm \eta}).
\end{align}
 In this approach, we could also include unknown  parameters of the prior density over ${\bm \eta}$, i.e., $p(\y|{\bm \eta}_\ell,{\bm \eta}_p) = \int_{\bm \Theta} \ell(\y|{\bm \theta},{\bm \eta}_\ell) g({\bm \theta}|{\bm \eta}_p) d{\bm \theta}$.
 
\item {\bf Type-3 --- Nested models:} Nested models are models that belong to the same parametric family but, unlike in the previous scenario, the {\it complexity} of the model can change, i.e., the number of parameters  $|{\bm \Theta}_m|=D_{{\bm \theta}_m}$ is also unknown and must be inferred as well, jointly with the parameter ${\bm \theta}_m$. Namely, we have a sequence of likelihoods defined in an increasing dimensional space, such as $\ell({\bf y}|\theta_1,\mathcal{M}_1)$, $\ell({\bf y}|\theta_1,\theta_2,\mathcal{M}_2)$, $\ell({\bf y}|\theta_1,\theta_2,\theta_3,\mathcal{M}_3)$, etc. Some examples of this framework are:  variable selection, order selection  (in polynomial regression or ARMA models etc.),  clustering (when the number of clusters are unknown) and  dimension reduction problems, to  name a few \citep{bishop2006pattern}. 
\end{itemize}
%

\section{Use of vague priors and/or improper priors in Level-2}\label{FernandoMeMata}

For simplicity, hereafter, whenever we focus on a single although arbitrary model $\mathcal{M}_m$, we skip the dependence on $\mathcal{M}_m$ in the notation. For instance, we denote  the posterior density as $\bar{\pi}({\bm \theta}|\y)$ and  the marginal likelihood as $Z=p(\y)$. Thus, we write 
\begin{equation}\label{MarginalLikelihood2}
Z=\int_{\bm \Theta} \pi({\bm \theta}|\y) d{\bm \theta}= \int_{\bm \Theta} \ell(\y|{\bm \theta}) g({\bm \theta}) d{\bm \theta}.
\end{equation}
We can see clearly that $Z$ is an average of likelihood values $ \ell(\y|{\bm \theta})$, weighted according to the prior pdf $g({\bm \theta})$.

\subsection{Diffuse/vague priors are informative for model selection}\label{vaguePriorSects}
 If the support ${\bm \Theta}$ is unbounded and additional information is not available, one can employ a so-called {\it vague} prior density, i.e., a density with probability mass spread in all the state space, with a great scale parameter.
This kind of prior has different names such as {\it diffuse}, {\it  flat}, etc.  Let us consider now an illustrative example about the impact on the inference using a vague prior.


\subsubsection*{Illustrative example}

Let us assume a likelihood function that is integrable in every subset of an unbounded ${\bm \Theta}$, that is, for all  $A\subseteq {\bm \Theta}$, $ \int_{A \in \bm{\Theta}} \ell(\y|{\bm \theta})d{\bm \theta} < \infty$.
In particular, when $A={\bm \Theta}$, the integral corresponds to the ``area below'' the likelihood function,
	\begin{align}
		S =  \int_{{\bm \Theta}} \ell(\y|{\bm \theta})d{\bm \theta} < \infty.
\end{align}
Hence, in this scenario, the normalized likelihood is a proper pdf on ${\bm \Theta}$. 
Then, we consider a uniform and proper prior defined on the hyper-volume $B$, i.e.,
$$g({\bm \theta}) = \frac{1}{|B|} \bm{1}_{B}({\bm \theta}),
$$
where $|B|$ represents the volume of $B$. Hence, the posterior pdf is 
\begin{align}
	\post({\bm \theta}|\y) = \frac{\ell(\y|{\bm \theta})\bm{1}_{B}({\bm \theta})}{\int_B \ell(\y|{\bm \theta})d{\bm \theta}},
\end{align}
which is the normalized likelihood restricted to the set $B$. We discuss what happens  in  Level-1 and  Level-2 as $|B| \to \infty$.
\begin{itemize}
\item {\it  Level-1:} as we increase the volume of $B$, more and more mass of the likelihood is considered. As $|B| \rightarrow \infty$, we have that $\post({\bm \theta}|\y)$ becomes closer and closer to   
\begin{align}
	\post^*({\bm \theta}|\y) = \frac{\ell(\y|{\bm \theta})}{\int_{{\bm \Theta}} \ell(\y|{\bm \theta})d{\bm \theta}}= \frac{\ell(\y|{\bm \theta})}{S}.
\end{align}
Namely,  in the limit where $B={\bm \Theta}$, the prior $g({\bm \theta})$ becomes equivalent to an improper uniform prior on ${\bm \theta}$.  The posterior $\post^*({\bm \theta}|\y)$ contains only the information provided by the likelihood function, and is not affected or distorted by the prior. Hence, a vague/diffuse prior (or a  uniform improper prior, which is its maximal expression) can be employed for expressing the absence of additional information in the choice of the prior (at Level-1 of inference).


\item {\it Level-2:} we focus now on the marginal likelihood $Z$ which, in this case, is given by
	\begin{align}\label{SuperImpZlucaParadox}
		Z = \frac{\int_B \ell(\y|{\bm \theta})d{\bm \theta}}{|B|}.
	\end{align}
Let us consider increasing $B$ until we cover all parameter space, i.e.,
$$
|B| \to \infty, \quad \text{but} \quad \int_B \ell(\y|{\bm \theta})d{\bm \theta} \to S,
$$
Hence,
\begin{align}
	\lim_{|B| \to \infty}Z = 0.
\end{align}
We see that the marginal likelihood of a model with a increasingly-diffuse prior becomes zero.
Hence,  in model selection (Level-2), actually vague/diffuse priors are highly informative, in the sense that, (if $S$ is finite) an increasingly diffuse prior penalizes more and more the considered model, so that their use has a clear impact to the results of the model selection. 
\end{itemize}
Hence, we can highlight two conclusions.

{\Remark In the Level-1 of inference, if $S =  \int_{{\bm \Theta}} \ell(\y|{\bm \theta})d{\bm \theta}$ is finite, we can use vague prior as non-informative (or weakly-informative) choice, since the idea is to perform the minimum possible perturbation to the likelihood function and, as a consequence, a minimum impact to the inference of ${\bm \theta}$ (and, generally, we can asymptotically recover some frequentist results). 
}

{\Remark In Level-2 inference, the choice of a diffuse/vague prior is actually very informative. For instance, if $S= \int_{{\bm \Theta}} \ell(\y|{\bm \theta})d{\bm \theta}$ is finite, diffuse priors tend to produce smaller values of the marginal likelihood $Z$    \citep{cameron2014recursive,Bernardo94}.  
}
\newline
\newline
Hence, if $S<\infty$, a good model can display a low value of $Z$ only because we choose a prior that is very spread out. Conversely, a worse model can display a bigger value of $Z$ due to choosing a concentrated prior \citep{Bernardo94,mackay2003information,r2019marginal,LlorenteREV19}.


\subsection{Improper priors: forbidden for computing the evidence $Z$}

Let us consider again that the domain ${\bm \Theta}$ is unbounded. An improper prior is such that 
\begin{align}
\int_{\bm \Theta} g({\bm \theta}) d{\bm \theta}=\infty.
\end{align}
Note that in this case, the prior $g(\boldsymbol{\theta})=c \cdot h(\boldsymbol{\theta})$ (where $\int_{\bm \Theta} h({\bm \theta}) d{\bm \theta}=\infty$)   is not completely specified, since $h$ cannot be normalized, i.e.,  the normalization constant  $c$ does not exist, and as a consequence, the constant $c$ is arbitrary. Let us assume that, however,  
\begin{align}\label{EqZlh}
 \int_{\bm \Theta} \ell(\y|{\bm \theta}) h({\bm \theta}) d{\bm \theta}= Z_{\ell \times h} <\infty
\end{align}
is finite.  We call $Z_{\ell \times h}$ as {\it fake evidence}. In this case, trying to computing the Bayesian evidence $Z$, we obtian 
\begin{align}
Z&= \int_{\bm \Theta} \ell(\y|{\bm \theta}) g({\bm \theta}) d{\bm \theta}, \nonumber   \\
&= c \int_{\bm \Theta} \ell(\y|{\bm \theta}) h({\bm \theta}) d{\bm \theta},  \\
&= c \ Z_{\ell \times h}, \nonumber  
\end{align}
i. e., the marginal likelihood $Z$  is also not completely specified due to $c>0$ is unknown/arbitrary. Then, we can remark below:

{\Remark Improper priors can not be used for computing marginal likelihood $Z$.  Thus, generally,  improper priors are not allowed for model selection ({\bf Level-2} of inference). However, we will see  that there is an exception for Type-2 of model selection. }
\newline
\newline
 On the other hand, the use of an improper prior, i.e., is allowed for {\bf Level-1} inference when $Z_{\ell \times h}$ in Eq.  \eqref{EqZlh} is finite. Indeed, in this case,
 the corresponding posterior is still proper,
 \begin{align*}
 \post({\bm \theta}|\y)&= \frac{\ell(\y|{\bm \theta}) g({\bm \theta})}{\int_{\bm \Theta} \ell(\y|{\bm \theta}) g({\bm \theta}) d{\bm \theta}}, \\
&= \frac{\cancel{c} \ \ell(\y|{\bm \theta}) h({\bm \theta})}{\cancel{c}\int_{\bm \Theta} \ell(\y|{\bm \theta}) h({\bm \theta}) d{\bm \theta}}, \\
&= \frac{\ell(\y|{\bm \theta}) h({\bm \theta})}{Z_{\ell \times h}},
 \end{align*}
  since $Z_{\ell \times h}=\int_{\bm \Theta} \ell(\y|{\bm \theta}) h({\bm \theta}) d{\bm \theta}  <\infty$ is finite.
   
 {\Remark  Improper priors are allowed in {\bf Level-1} of inference if the fake evidence  $Z_{\ell \times h}$ is finite, i.e.,  $Z_{\ell \times h}<\infty$ (since the corresponding posteriors are still proper).  }

\subsection{Uniform improper prior and the area under the likelihood}

An extreme case of vague prior and the simplest example if improper prior is {\it the uniform improper prior}, i.e.,  $g({\bm \theta})\propto h({\bm \theta})=1$ for all ${\bm \theta}$ in the unbounded support ${\bm \Theta}$. It is often employed for expressing the absence of a-priori information in the {\bf Level-1} of inference when the {\it area under the likelihood} ($S$) is finite, i.e.,
\begin{align}\label{AUL}
S({\bf y}){= Z_{\ell \times 1}}=\int_{\bm \Theta} \ell(\y|{\bm \theta})  d{\bm \theta} <\infty.
\end{align}
Indeed, in this case the unnormalized posterior $\pi({\bm \theta}|\y)=\ell(\y|{\bm \theta})$ can be normalized as 
\begin{align}
\post({\bm \theta}|\y)=\frac{1}{S({\bf y})}\pi({\bm \theta}|\y)=\frac{1}{S({\bf y})} \ell(\y|{\bm \theta}).
\end{align}
Hence, we need $S({\bf y})<\infty$, in order to be able to use improper uniform prior in Level-1 of inference. 
Note also that $S({\bf y})>0$ since $\ell(\y|{\bm \theta})$ must be positive in a a region with non-zero measure. 


 {\Remark The quantity in Eq. \eqref{AUL},  i.e., $S({\bf y})$, cannot be interpreted as an evidence $Z=p({\bf y})$, since the values of the likelihood $\ell(\y|{\bm \theta})$ are not weighted (by a prior density). It is a special case of fake evidence $Z_{\ell \times h}$ when $h({\bm \theta})=1$. }
\newline
\newline
 However, we show and remark that the area under the likelihood $S$ or, the fake evidence $Z_{\ell \times h}$, can be still useful in Type-2 of model selection \ref{TypeModSection}. 
 

\section{Key observations}\label{KEYPOINT}

 Tuning the parameters (or hyper-parameters) in a family of models is a special scenario of model selection, i.e., Type-2 described in Section \ref{TypeModSection}. We will see that, in this specific case, we can use improper priors since the fake evidences $Z_{\ell \times h}$ are meaningful in some sense. 
\subsection{Comparing models which differ for the chosen parameters} 

For simplicity, let us  consider  two models which differ only for the tuning of the parameters ${\bm \eta}_\ell$ of the likelihood function (induced by the observation model){, and for ${\bm \eta}_{p}$ some parameters of the prior.} More specifically, let say that we have $\ell(\y|{\bm \theta},{\bm \eta}_{\ell,1})$ and $ \ell(\y|{\bm \theta},{\bm \eta}_{\ell,2})$, considering the use of {\it the same type} of improper prior for both models {but with different hyper-parameters ${\bm \eta}_{p,1}$ and ${\bm \eta}_{p,2}$, i.e., $g({\bm \theta}|{\bm \eta}_{p,i})=c \cdot h({\bm \theta}|{\bm \eta}_{p,i})$ for $i=1,2$}, the Bayes factor is
\begin{align*}
\mbox{BF}_{12}=\frac{Z_1}{Z_{2}} &=\frac{\int_{\bm \Theta} \ell(\y|{\bm \theta},{\bm \eta}_{\ell,1}) g({\bm \theta}{|{\bm \eta}_{p,1}}) d{\bm \theta}}{\int_{\bm \Theta} \ell(\y|{\bm \theta},{\bm \eta}_{\ell,2}) g({\bm \theta}{|{\bm \eta}_{p,2}}) d{\bm \theta}}, \\
&=\frac{\cancel{c}\int_{\bm \Theta} \ell(\y|{\bm \theta},{\bm \eta}_{\ell,1}) h({\bm \theta}{|{\bm \eta}_{p,1}}) d{\bm \theta}}{\cancel{c} \int_{\bm \Theta} \ell(\y|{\bm \theta},{\bm \eta}_{\ell,2}) h({\bm \theta}{|{\bm \eta}_{p,2}}) d{\bm \theta}}, \\
&=\frac{Z_{\ell_1\times h_1}}{Z_{\ell_2\times h_2}}, 
\end{align*} 
 i.e., the ratio of marginal likelihoods $\frac{Z_1}{Z_{2}} $ is well-defined in this scenario, equal to the ratio fo fake evidences $\frac{Z_{\ell_1\times h}}{Z_{\ell_2\times h}}$. {Note that the constant $c$ of the prior must disappear and cannot be included in the vectors ${\bm \eta}_{p,i}$.} 
 \newline
 More generally, considering $M$ possible ${\bm \eta}_{\ell,m}$, $m=1,...,M$, vector of parameters  (i.e., $M$ possible models) and using the same  {\it the same type} improper prior for all the models, $g({\bm \theta}|{{\bm \eta}_{p,i}})=c \cdot h({\bm \theta}|{{\bm \eta}_{p,i}})$, we could apply a Bayesian model averaging \cite{BMA99} with the following normalized weights:
 $$
 \bar{w}_m=\frac{Z_m}{\sum_{i=1}^M Z_i} =\frac{\cancel{c} Z_{\ell_m\times h_m}}{\cancel{c}\sum_{i=1}^M Z_{\ell_i\times h_i}}, \quad m=1,...,M, 
 $$ 
 that are again well-defined, since the arbitrary value $c$ is cancelled out.  In the simplest case of a uniform improper prior, we have $\bar{w}_m=\frac{S_m}{\sum_{i=1}^M S_i}$,  $m=1,...,M$.
In other words, the improper prior here can be employed since it is shared by all the models.  

 {\Remark In the scenario of tuning some parameters ${\bm \eta}$ of the observation model, the use of unique improper prior over ${\bm \theta}$ (the same prior for all models) is allowed, and 
the fake evidence $Z_{\ell_m\times h}$ can be employed for comparing models.}
\newline
\newline
Hence, the fake evidence $Z_{\ell_m\times h_m}$  can be employed in Type-2 of model selection.
 
 {\Remark  In this sense, the statement ``improper priors are not allowed for model selection'' is technically wrong \cite{LlorenteSafePriors}. A more correct statement is ``improper priors are not allowed for computation of the Bayesian evidence $Z$'', but they can be used in Type-2 of model selection computing the fake evidence $Z_{\ell_m\times h_m}$.}
 \newline
\newline
{In Type-3 of model selection, i.e., the nested model scenario, an improper prior is allowed only for the first variable/parameter, common and shared to all the nested models. }
\subsection{On the area under the likelihood $S$}
Let us consider now the case $h({\bm \theta})=1$, i.e.,  $Z_{\ell_m\times 1}=S$ with $S=\int_{\Theta} \ell(\y|{\bm \theta}) d{\bm \theta}<\infty$.  { With respect to the use of $h({\bm \theta})=1$, with respect to the fake evidence, $Z_{\ell \times h}$, with the area under the likelihood, $S$:

  {\Remark  Unlike the fake evidence $Z_{\ell \times h}$ for a generic non-uniform $h({\bm \theta})$, the quantity $S=Z_{\ell \times 1}$ in Eq. \eqref{AUL} can be used only for tuning parameters of the likelihood function ${\bm \eta}_{\ell}$, since the only possible parameter to add in $h({\bm \theta})=1$ is a multiplicative constant that cannot be included in ${\bm \eta}_p$.
 }
 \newline
 \newline
 Moreover, $S$ cannot be recovered as an asymptotic special case using  a proper prior density (as one could expect from other results in Level-1 of inference):
}

{\Remark The value of the area under the likelihood $S$ cannot be obtained starting with a diffuse prior and then increase its scale parameter to infinity (as usually done in Level-1 of inference). Namely, let us consider $S<\infty$. Note that $S>0$ since $\ell(\y|{\bm \theta})>0$ in a a region with non-zero measure. As the  scale of the diffuse prior grows, $Z\rightarrow 0$, hence $Z \nrightarrow S$.   }

{
\subsection{On the empirical Bayes and profile likelihood approaches}

For simplicity, in this section, we consider $h({\bm \theta})=1$ but all comments in this section are valid for the more general case.  We can extend the previous considerations in order to compare a {\it continuous} of models, i.e., infinite models belonging to the same parametric family. In this sense, find the best model is equivalent} to obtain a point-wise estimator of ${\bm \eta}$ as
\begin{align}
{\bm \eta}^*=\arg\max S(\y|{\bm \eta})=\arg\max   \int_{\bm \Theta} \ell(\y|{\bm \theta},{\bm \eta}) d{\bm \theta}.
\end{align}
With respect to ${\bm \eta}$, if we are just interested in a pointwise estimator ${\bm \eta}^*$, we are basically employing a frequentist approach over ${\bm \eta}$ { but after integrating out ${\bm \theta}$ from the likelihood. Hence, it can be interpreted as a combination of  Bayesian-frequentist approaches. 
 Indeed, it differs from the classical maximization of the complete likelihood function $\ell(\y|{\bm \theta},{\bm \eta})$ (see Sections \ref{AULsect_2} and \ref{NumSect0}), that is a completely frequentist strategy. This is called {\it empirical Bayes} approach, and can be also employed to select unknown parameters of the prior. }
 \newline
  { A similar and completely frequentist approach, is the {\it profile likelihood} of a parameter of interest, which is defined as 
$$
L_p({\bm \eta})=\ell(\y|{\bm \theta}_{\bm \eta}^*,{\bm \eta})=\max _{{\bm \theta}}  \ell(\y|{\bm \theta},{\bm \eta}), \qquad  {\bm \theta}_{\bm \eta}^*=\arg\max _{{\bm \theta}}  \ell(\y|{\bm \theta},{\bm \eta}).
$$
The idea is to study this function to derive confidence regions  and other statistical information (similarly with a posterior density). 
A normalized version can be obtained by dividing this expression by the  complete maximum likelihood value, i.e., $
{\widetilde L}_p({\bm \eta})=\frac{\max _{{\bm \theta}}  \ell(\y|{\bm \theta},{\bm \eta})}{\max _{{\bm \theta},{\bm \eta}}  \ell(\y|{\bm \theta},{\bm \eta})}$. However, in the profile likelihood there is not any integration and there is not any explicit, implicit prior density involved. Some advantages of integrating out ${\bm \theta}$ are shown in Section Sections \ref{AULsect_2} and \ref{NumSect0}.}

{
\subsection{Full-Bayesian solution with a double improper (uniform) prior}
}

{ Again, here we consider $h({\bm \theta})=1$ but all considerations can be extended to a generic $h({\bm \theta})$.
As an alternative approach,} if $S_Z=\int_{\bm \Theta} S(\y|{\bm \eta}) d {\bm \eta}< \infty$, we  {could} consider {again an} improper uniform prior over ${\bm \eta}$ and the marginal posterior would be $p({\bm \eta}|{\bf y})=\frac{1}{S_Z} S(\y|{\bm \eta})$ { (other type of improper prior can be also employed).  Thus, the so-called {\it full-Bayesian solution} could be obtained considering improper priors twice. In this case, the full-Bayesian solution can be performed following the steps below:
 \begin{enumerate}
 \item Draw ${\bm \eta}_1,...,{\bm \eta}_R$ from $p({\bm \eta}|{\bf y})=\frac{1}{S_Z} S(\y|{\bm \eta})$.
 \item Draw ${\bm \theta}_{r,1},...,{\bm \theta}_{r,N}$ from $p({\bm \theta}|{\bf y},{\bm \eta}_r)=\frac{\ell(\y|{\bm \theta},{\bm \eta}_r)}{S(\y|{\bm \eta}_r)}\propto \ell(\y|{\bm \theta},{\bm \eta}_r) $, for $r=1,...,R$. 
 \end{enumerate}
 The outputs are the $NR$ samples, $\{{\bm \theta}_{r,n}\}$, and $R$ samples $\{{\bm \eta}_{r}\}$ for $r=1,...,R$ and $n=1,...,N$. With this procedure, the model parameter ${\bm \eta}$ is marginalized out. Another way to interpret this procedure is as a continuous Bayesian model averaging, i.e., consider a resulting model expressed as  
 \begin{align}\label{SuperAveModel}
 {\bar \ell}(\y|{\bm \theta})=\int \ell(\y|{\bm \theta},{\bm \eta}) p({\bm \eta}|{\bf y})d {\bm \eta},
  \end{align}
  where we are weighting the different models, $\ell(\y|{\bm \theta},{\bm \eta})$, according to the marginal posterior $p({\bm \eta}|{\bf y})$. In the  integral in Eq. \eqref{SuperAveModel},  this marginal posterior $p({\bm \eta}|{\bf y})$ is employed as an informative prior over ${\bm \eta}$, but it is an {\it objective prior} built after looking the data \cite{LlorenteSafePriors}.
Moreover recall that, if $S$ and $S_Z$ are both finite, the use of the improper priors is allowed twice in this procedure.}
Table \ref{TablaResumen} provides a summary of the main concepts. In the next section, we check and confirm the previous statements in a Bayesian regression setting.

\begin{table}[!h]	
	 \caption{Summary of the main concepts.}\label{TablaResumen}
	 \vspace{-0.2cm}
	 \footnotesize
	\begin{center}
		\begin{tabular}{|c|l|l|} 
		\hline 
		{\bf Prior densities }	& {\bf Level-1 of inference} & {\bf  Level-2 of inference}    \\ 
			\hline 
			\hline 
			 &  & \\
		Diffuse/vague priors	 &  weakly informative. & informative;   \\
		 & & If $S<\infty$, then $Z \rightarrow 0$ as the prior\\
		 & &  becomes more diffuse. \\
		 \hline
		 \hline
		 & & \\
		 & non-informative; & They are not allowed for computing $Z$; \\
		Improper priors	 & If $S<\infty$, they can be used. & If $S<\infty$, they can be used  for Type-2 \\
		& to make inference  on ${\bm \theta}$. & of model selection.\\
		& & \\
		\hline
		 \hline
			 &  We asymptotically obtain the same results  &  \\
		From diffuse $\rightarrow$	  & using an improper uniform prior;   & If $S<\infty$ , then $Z \rightarrow 0$,\\
		 to improper uniform & Generally, we recover  &   hence $Z  \nrightarrow S$ {(since $S>0$)}. \\
		 & some frequentist results. &  \\
	\hline 
		\end{tabular}
	\end{center}
\end{table} 

{
\noindent
It is  also important to remark  that all the theoretical and practical considerations (in this work) are valid either when the marginal likelihood can be computed analytically, or when the marginal likelihood is intractable. In the latter scenario, we have just the additional computational problem of approximating the  marginal likelihood \cite{LlorenteREV19}.  
}

\section{Example of application to Bayesian regression models}\label{Chap6FirstSect}

In this section, we consider a generalized linear model for regression, considering $N$ data points and $M$ different non-linear bases, with $M<N$. We apply two types of priors to the vector of coefficients ${\bm \theta}$: an improper uniform prior and a Gaussian prior. In both cases, we give a complete Bayesian analysis and try to design a Level-2 of inference in order to infer a vector of parameters of the bases ${\bm \alpha}$, the noise power $\sigma_e^2$ and the rest of nuisance parameters. Firstly, the goal of this section is to show some applicative examples. Secondly, the goal is confirm some important statements provided above, from a more practical point of view. Finally, this section gives the theoretical support for the numerical example in Section \ref{NumSect}. 


\subsection{Problem statement}

Let us consider the dataset $\{{\bf x}_n,y_n\}_{n=1}^N$, where ${\bf x}_n=[x_{n,1},\ldots,x_{n,d_X}]\in \mathcal{X}  \subseteq \mathbb{R}^{d_X}$ represents the inputs, $y_n \in  \mathbb{R}$, denotes the outputs. The vector of outputs is then ${\bf y}=[y_1,\ldots,y_N]^{\top} \in  \mathbb{R}^N$.
We consider the following observation probabilistic model which link the vectors ${\bf x}$ and ${\bf y}$,
\begin{eqnarray}
y&=&f({\bf x})+ e,  \quad e\sim \mathcal{N}(e|0,\sigma_e^2). \nonumber 
\end{eqnarray}
We assume that the underlying function can have the following parametric form,
\begin{equation}
f({\bf x})
=\sum_{m=1}^M \phi_m({\bf x},{\bm \alpha}) \theta_m ={\bm \phi}({\bf x},{\bm \alpha})^{\top} {\bm \theta}, \qquad M \ { \leq } \ N,
\end{equation}
 where $\phi_m({\bf x},{\bm \alpha}): \mathcal{X}  \times \Omega \rightarrow \mathbb{R}$ is the $m$-th nonlinear function where ${\bm \alpha}\in \Omega \subseteq \mathbb{R}^{d_\theta}$ represents a vector of parameters, that the user have to tune \cite{bishop2006pattern}. Defining the vectors
\begin{eqnarray}
{\bm \phi}({\bf x},{\bm \alpha})&=&[\phi_1({\bf x},{\bm \alpha}),\phi_2({\bf x},{\bm \alpha}),\ldots,\phi_M({\bf x},{\bm \alpha}) ]^{\top},  \\
{\bm \theta}&=&[\theta_1,\theta_2,\ldots,\theta_M ]^{\top},  
\end{eqnarray}
then we can rewrite the model above as 
\begin{align}
y=\sum_{m=1}^M \phi_m({\bf x},{\bm \alpha}) \theta_m+e=\underbrace{ {\bm \phi}({\bf x},{\bm \alpha})^{\top} }_{1\times M} \underbrace{{\bm \theta}}_{M\times 1}+e.
\end{align}
Hereafter, for simplicity,  we will remove the dependence of ${\bm \alpha}$, so that $ \phi_m({\bf x})= \phi_m({\bf x},{\bm \alpha})$ and ${\bm \phi}({\bf x})={\bm \phi}({\bf x},{\bm \alpha})$. For instance, we will write simply $f({\bf x})={\bm \phi}({\bf x})^{\top} {\bm \theta}$.
\newline 
\newline
{\bf Vectorial form.} The model above can be written in a vectorial form as 
\begin{eqnarray}
\underbrace{{\bf y}}_{N\times 1}&=&\underbrace{{\bf f}}_{N\times 1}+ \underbrace{{\bf e}}_{N \times 1} \label{estoMeint1}\\
&=&\underbrace{{\bm \Phi}}_{N\times M} \underbrace{{\bm \theta}}_{M\times 1}+ \underbrace{{\bf e}}_{N \times 1},\label{estoMeint2}
\end{eqnarray}
where  we have denoted ${\bf f}={\bm \Phi} {\bm \theta}$, ${\bf e}\sim \mathcal{N}({\bf 0},\sigma_e^2 {\bf I}_N)$ and  we have defined $N\times M$ {\it design matrix} ${\bm \Phi}=[{\bm \phi}({\bf x}_1),\ldots., {\bm \phi}({\bf x}_N)]^{\top}$ (that is rectangular, in general), i.e., 
\begin{eqnarray}
\label{phiMAT_capantes}
\underbrace{{\bm \Phi}}_{N\times M} =
\begin{bmatrix}
\phi_1({\bf x}_1) &\phi_2({\bf x}_1) &  \hdots  &  \phi_M({\bf x}_1) \\
\phi_1({\bf x}_2) & \phi_2({\bf x}_2) & \hdots & \phi_M({\bf x}_2) \\
\vdots \\
\phi_1({\bf x}_N) & \phi_2({\bf x}_N) &\hdots & \phi_M({\bf x}_N) 
\end{bmatrix}.
\end{eqnarray}
{ Note that, in this section, all the considerations are valid for  $M\leq N$. The case $M=N$ is also included: the resulting regression method becomes non-parametric. The scenario $M>N$, since requires specific observations and analysis,  is not considered here.  }



\subsection{Likelihood  function}
The observation model above induce a likelihood function with respect to (w.r.t.) the coefficients ${\bm \theta}$, that is
\begin{align}\label{likefunctionEq}
\ell({\bf y}| {\bm \theta})=\ell({\bf y}| {\bm \theta},{\bm \Phi},{\bm \alpha},\sigma_e^2)&=\left(2 \pi \sigma_e^{2}\right)^{-\frac{N}{2}} \exp \left(-\frac{({\bf y}-{\bm \Phi} {\bm \theta})^{\top}({\bf y}-{\bm \Phi} {\bm \theta})}{2 \sigma_e^{2}}\right) \nonumber \\
&=\left(2 \pi \sigma_e^{2}\right)^{-\frac{N}{2}} \exp\left(-\frac{||{\bf y}-{\bm \Phi} {\bm \theta}||^2}{2 \sigma_e^{2}}\right)  \nonumber \\
& = \mathcal{N}({\bf y}| {\bm \Phi} {\bm \theta}, \sigma_e^2 {\bf I}_N),
\end{align}
where ${\bf I}_N$ is the $N \times N$ identity matrix.
Clearly, a more complete notation would be $\ell({\bf y}| {\bm \theta}, {\bm \Phi},{\bm \alpha}, \sigma_e)$. However, we first focus on the coefficients ${\bm \theta}$ and consider, in this first stage, 
${\bm \Phi}$ and the nonlinear bases are chosen in advance. The parameters ${\bm \alpha}$ and $\sigma_e$ should be tuned and decided by the user. Then, the complete vector of hyper-parameters, denoted as ${\bm \lambda}$, is formed by ${\bm \alpha}$ and $\sigma_e$, i.e., we have ${\bm \lambda}=[{\bm \alpha},\sigma_e]$ \cite{bishop2006pattern,bayesian_choice}. { Fixing the bases  and ${\bm \alpha}$ the classical maximum likelihood approach, i.e.,
 \begin{align}\label{MLestAQUI}
\left[\widehat{{\bm \theta}}_{\texttt{ML}},\widehat{\sigma}_{e,\texttt{ML}}^2\right]=\arg \max_{{\bm \theta},\sigma_e^2}\ell({\bf y}| {\bm \theta},\sigma_e^2),
 \end{align}  
 provides the following estimators (that can be obtained analytically in this case):
  \begin{align}\label{WestML0}
\widehat{ {\bm \theta}}_{\texttt{ML}}=({\bm \Phi}^{\top}{\bm \Phi})^{-1} {\bm \Phi}^{\top} {\bf y}, \qquad \widehat{\sigma}_{e,\texttt{ML}}^2=\frac{1}{N} ||{\bf y}-{\bf \widehat{f}}||^2.
 \end{align}
 Note that $\widehat{\sigma}_{e,\texttt{ML}}^2$ is a consistent, but  biased estimator.
}

\subsection{Uniform improper prior over ${\bm \theta}$}

In this section, we assume a uniform {\it improper} prior density over the weights ${\bm \theta}$, i.e., $g({\bm \theta}) \propto 1$ for all ${\bm \theta}$.

\subsubsection{Posterior of the coefficients ${\bm \theta}$}
\label{PostTheta1}
 Therefore, the posterior pdf of the coefficient ${\bm \theta}$ is 
\begin{align}
 \post({\bm \theta}| {\bf y})=\frac{ \ell({\bf y}| {\bm \theta})g({\bm \theta})}{p({\bf y})} &\propto \ell({\bf y}| {\bm \theta})g({\bm \theta}), \nonumber\\
&\propto \ell({\bf y}| {\bm \theta}),  \nonumber \\
&\propto \left(2 \pi \sigma_e^{2}\right)^{-\frac{N}{2}} \exp\left(-\frac{||{\bf y}-{\bm \Phi} {\bm \theta}||^2}{2 \sigma_e^{2}}\right),
\end{align}
i.e., proportional to the likelihood function $\ell({\bf y}| {\bm \theta})$ (we have use $g({\bm \theta}) \propto 1$). 
After some algebra and rearrangements (in order to express the formula as a Gaussian density with respect to ${\bm \theta}$ instead of ${\bf y}$), we can express $\post({\bm \theta}| {\bf y})$ as a Gaussian distribution with mean vector ${\bm \mu}_{\theta|y}$ and covariance matrix ${\bm \Sigma}_{\theta|y}$ \cite{Fitzgerald96_bo}, i.e.,
\begin{align}
 \post({\bm \theta}| {\bf y})= \mathcal{N}({\bm \theta}|{\bm \mu}_{\theta|y},{\bm \Sigma}_{\theta|y}),
\end{align}
with 
\begin{equation}\label{West}
{\bm \mu}_{\theta|y}=\widehat{{\bm \theta}}=(\underbrace{{\bm \Phi}^{\top}{\bm \Phi}}_{M\times M})^{-1} {\bm \Phi}^{\top} {\bf y},
\end{equation}
and 
\begin{equation}
{\bm \Sigma}_{\theta|y}=\left(\frac{1}{\sigma_e^2}{\bm \Phi}^{\top}{\bm \Phi}\right)^{-1} =\sigma_e^2 \left({\bm \Phi}^{\top}{\bm \Phi}\right)^{-1}. 
\end{equation}

{\Remark Note that, using a uniform {\it improper} prior, the posterior density $ \post({\bm \theta}| {\bf y})$  
over the coefficient vector resembles the ``frequentist'' sampling distribution described in Appendix \ref{App1}, being both Gaussian with the same mean and covariance matrix, although they have a complete statistical different meaning (see App. \ref{App1}). Moreover, $\widehat{{\bm \theta}}$ coincides with the maximum likelihood estimator, i.e., $\widehat{{\bm \theta}}=\widehat{{\bm \theta}}_{\texttt{ML}}$ (see App. \ref{App1} and Eq. \eqref{WestML0}).
}
\subsubsection{Posteriors of the function $f(\x)$ and vector ${\bf f}$}

 Let us recall that the assumed model is $f(\x)={\bm \phi}({\bf x})^{\top} {\bm \theta}$ and ${\bm \theta} \sim \post({\bm \theta}| {\bf y})= \mathcal{N}({\bm \theta}|{\bm \mu}_{\theta|y},{\bm \Sigma}_{\theta|y})$, after seeing the data.  Hence, given a fixed $\x$,  
the hidden function $f(\x)$ is a random variable with a Gaussian posterior density, 
\begin{align}
p(f(\x) | {\bf y})= \mathcal{N}(f(\x)|\mu_{f|y}(\x),\sigma_{f|y}^2(\x)),
\end{align}
with  mean at ${\bf x}$,
\begin{align}
\mu_{f|y}(\x)=\widehat{f}(\x)&={\bm \phi}({\bf x})^{\top} \widehat{{\bm \theta}}  \nonumber \\
&={\bm \phi}({\bf x})^{\top}({\bm \Phi}^{\top}{\bm \Phi})^{-1} {\bm \Phi}^{\top} {\bf y}, \label{fest_bases}
\end{align}
and variance
\begin{equation}
\sigma_{f|y}^2(\x)=\sigma_e^2 {\bm \phi}({\bf x})^{\top}\left({\bm \Phi}^{\top}{\bm \Phi}\right)^{-1} {\bm \phi}({\bf x}),
\end{equation}
where we have considered the previous results regarding the posterior over the coefficients ${\bm \theta}$.
We remark that the regression function is the mean solution, i.e., $\widehat{f}(\x)=\mu_{f|y}(\x)={\bm \phi}({\bf x})^{\top}({\bm \Phi}^{\top}{\bm \Phi})^{-1} {\bm \Phi}^{\top} {\bf y}$ \cite{Fitzgerald96_bo}.
\newline
\newline
{\bf Posterior of the vector ${\bf f}$. }
In the smoothing, considering only estimations at the input features, i.e., ${\bf f}={\bm \Phi}{\bm \theta}$. Since   ${\bm \theta} \sim p({\bm \theta}| {\bf y})= \mathcal{N}({\bm \theta}|{\bm \mu}_{\theta|y},{\bm \Sigma}_{\theta|y})$, after seeing the data, the posterior of the vector of ${\bf f}$ is 
 \begin{align}
p({\bf f} | {\bf y})= \mathcal{N}({\bf f}|{\bm \mu}_{f|y},{\bm \Sigma}_{f|y}),
\end{align}
with the mean vector
\begin{align}\label{Fsmoothing}
\underbrace{{\bm \mu}_{f|y}}_{N\times 1}=\widehat{{\bf f}}={\bm \Phi}\widehat{{\bm \theta}}={\bm \Phi}({\bm \Phi}^{\top}{\bm \Phi})^{-1} {\bm \Phi}^{\top} {\bf y},
\end{align}
and with the covariance matrix 
\begin{align}
\underbrace{{\bm \Sigma}_{f|y}}_{N\times N}=\sigma_e^2 {\bm \Phi}\left({\bm \Phi}^{\top}{\bm \Phi}\right)^{-1} {\bm \Phi}^{\top}.
\end{align}

%
%

\subsubsection{Area under the likelihood ($S$)}\label{AULsect_2}  

As we have remarked in the previous sections,  sice we are using improper priors, we can just compute the {\it area under the likelihood} ($S$), instead of a well-defined marginal likelihood. The $S$ can be useful in certain scenarios, for instance, performing  an empirical Bayes approach.  
Indeed, the marginal likelihood $Z=p(\y)$ is defined as the integral
$ p({\bf y})=\int_{\mathbb{R}^N} \ell({\bf y}| {\bm \theta})g({\bm \theta}) d{\bm \theta}$.
 However, since we are employing an improper prior $g({\bm \theta}) \propto 1$, the marginal likelihood is not perfectly determined (a multiplicative factor is undetermined). As a consequence, we can just compute $S$, i.e., $S({\bf y})=\int_{\mathbb{R}^N} \ell({\bf y}| {\bm \theta}) d{\bm \theta}$, 
where we miss the probability interpretation. Again, a more complete notation would be $S({\bf y}|{\bm \Phi},{\bm \alpha}, \sigma_e)=S({\bf y}|{\bm \Phi}, {\bm  \lambda})$. Here, we focus on the choice of the hyper-parameters ${\bm  \lambda}=[{\bm \alpha},\sigma_e]$ that we should be tuned.  Then,  we write
$$
 S({\bf y})=S({\bf y}|{\bm  \lambda})=\int_{\mathbb{R}^N} \ell({\bf y}| {\bm \theta},{\bm  \lambda}) d{\bm \theta}.
$$
It is possible to show that \cite{Fitzgerald96_bo}
\begin{align}
S({\bf y}|{\bm  \lambda}) =S({\bf y}|{\bm \alpha}, \sigma_e)
 &= \frac{\left(2 \pi \sigma_e^{2}\right)^{-\left(\frac{N-M}{2}\right)}}{\sqrt{\operatorname{det}\left[{\bm \Phi} ^{\top}{\bm \Phi}\right]}} \exp \left[-\left(\frac{\y^{\top}\y-\widehat{{\bf f}}^{\top}\widehat{{\bf f}}}{2 \sigma_e^{2}}\right)\right],  \\
 &= \frac{\left(2 \pi \sigma_e^{2}\right)^{-\left(\frac{N-M}{2}\right)}}{\sqrt{\operatorname{det}\left[{\bm \Phi} ^{\top}{\bm \Phi}\right]}} \exp \left[-\left(\frac{\y^{\top}\y-{\bf y}^{\top}{\bm \Phi}\left({\bm \Phi}^{\top}{\bm \Phi}\right)^{-1} {\bm \Phi}^{\top} \y}{2 \sigma_e^{2}}\right)\right],\label{AULaqui_Ex}
\end{align}
where we have used the equality $\widehat{{\bf f}}={\bm \Phi}\left({\bm \Phi}^{\top}{\bm \Phi}\right)^{-1} {\bm \Phi}^{\top} \y$.
Note that $\operatorname{det}\left[{\bm \Phi} ^{\top}{\bm \Phi}\right]>0$ since the matrix ${\bm \Phi}^{\top}{\bm \Phi}$ is symmetric, positive definite.
Moreover, $S({\bf y}|{\bm  \lambda})$ above just depends only on ${\bm  \lambda}$ ( we have integrated out ${\bm \theta}$). Above,  we have used the identity,
\begin{align}\label{superIdentityEqVamos}
\widehat{{\bf f}}^{\top}\widehat{{\bf f}}={\bf y}^{\top}{\bm \Phi}\left({\bm \Phi}^{\top}{\bm \Phi}\right)^{-1} {\bm \Phi}^{\top} \y.
\end{align}
Indeed, replacing the expression \eqref{Fsmoothing}, i.e., $\widehat{{\bf f}}={\bm \Phi}({\bm \Phi}^{\top}{\bm \Phi})^{-1} {\bm \Phi}^{\top} {\bf y}$, in the first side of the equation above, we have
\begin{eqnarray}
\widehat{{\bf f}}^{\top}\widehat{{\bf f}}&=&\Big({\bm \Phi}\left({\bm \Phi}^{\top}{\bm \Phi}\right)^{-1} {\bm \Phi}^{\top} \y\Big)^{\top}{\bm \Phi}\left({\bm \Phi}^{\top}{\bm \Phi}\right)^{-1} {\bm \Phi}^{\top} \y,  \nonumber \\
&=&( \y^{\top}{\bm \Phi}\left({\bm \Phi}^{\top}{\bm \Phi}\right)^{-1} {\bm \Phi}^{\top} ){\bm \Phi}\left({\bm \Phi}^{\top}{\bm \Phi}\right)^{-1} {\bm \Phi}^{\top} \y, \nonumber \\
&=& \y^{\top}{\bm \Phi}\cancel{\left({\bm \Phi}^{\top}{\bm \Phi}\right)^{-1}} \cancel{ {\bm \Phi}^{\top} {\bm \Phi}}\left({\bm \Phi}^{\top}{\bm \Phi}\right)^{-1} {\bm \Phi}^{\top} \y,  \nonumber\\
&=& \y^{\top}{\bm \Phi}\left({\bm \Phi}^{\top}{\bm \Phi}\right)^{-1} {\bm \Phi}^{\top} \y, \label{superIdentityEqVamos2}
\end{eqnarray}
Moreover, note also that
\begin{eqnarray}\label{superRes}
\widehat{{\bf f}}^{\top}\y
= \y^{\top}{\bm \Phi}\left({\bm \Phi}^{\top}{\bm \Phi}\right)^{-1} {\bm \Phi}^{\top} \y=\widehat{{\bf f}}^{\top}\widehat{{\bf f}},
\end{eqnarray}
Hence, we can write to an important equality:
\begin{align}\label{SuperRes}
||\y-\widehat{{\bf f}}||^2&=\y^{\top} \y+ \widehat{{\bf f}}^{\top}\widehat{{\bf f}}-2\widehat{{\bf f}}^{\top} \y, \nonumber \\
&=\y^{\top} \y+ \widehat{{\bf f}}^{\top}\widehat{{\bf f}}-2\widehat{{\bf f}}^{\top} \widehat{{\bf f}}, \nonumber \\
&=\y^{\top} \y-\widehat{{\bf f}}^{\top} \widehat{{\bf f}}, 
\end{align}
where we have used $\widehat{{\bf f}}^{\top}\y=\widehat{{\bf f}}^{\top}\widehat{{\bf f}}$ in Eq. \eqref{superRes}. 
Note that from \eqref{SuperRes}, we have always $\y^{\top} \y-\widehat{{\bf f}}^{\top} \widehat{{\bf f}}\geq 0$. Namely, the power of the outputs $\y^{\top} \y$ is always greater or equal than the power of the smoothing solution $\widehat{{\bf f}}^{\top} \widehat{{\bf f}}$, i.e., $\y^{\top} \y\geq \widehat{{\bf f}}^{\top} \widehat{{\bf f}}$. This is clearly due to the {\it denoising} effect \cite{Fitzgerald96_bo}.
\newline
\newline
Then, the $S$ can be rewritten in terms of the smoothing error $||\y-\widehat{{\bf f}}||^2$, i.e.,   
\begin{align}\label{aquiMLtheta}
S({\bf y}|{\bm  \lambda}) 
= \frac{\left(2 \pi \sigma_e^{2}\right)^{-\left(\frac{N-M}{2}\right)}}{\sqrt{\operatorname{det}\left[{\bm \Phi} ^{\top}{\bm \Phi}\right]}} \exp \left[-\left(\frac{||\y-\widehat{{\bf f}}||^2}{2 \sigma_e^{2}}\right)\right].
\end{align}
The negative log-$S$ is
\begin{align}\label{LogLikefirst}
\fbox{$C({\bm  \lambda})=-\log S({\bf y}|{\bm  \lambda}) =\frac{||\y-\widehat{{\bf f}}||^2}{2 \sigma_e^{2}} +\frac{N-M}{2} \log (2 \pi \sigma_e^{2})+\frac{1}{2}\log\operatorname{det}\left[{\bm \Phi} ^{\top}{\bm \Phi}\right].$}
\end{align}
We can try to minimize the cost function $C({\bm  \lambda})$ in \eqref{aquiMLtheta} with respect to ${\bm  \lambda}=[{\bm \alpha},\sigma_e]$. Alternatively, we can try to simplify the equation above. One possibility is shown below.
\newline
\newline
{\bf Estimator of the noise variance.}  If we keep fixed ${\bm \alpha}$ (and hence also the matrix ${\bm \Phi}$), it is possible to show that the conditional maximum value  w.r.t.  $\sigma_e$ (conditioned to ${\bm \alpha}$) is \cite{ATAIS} 
\begin{align}\label{NoiseEst1}
\widehat{\sigma}_e^2&=\frac{1}{N-M} ||\y-\widehat{{\bf f}}||^2, \\
&=\frac{1}{N-M} ||\y-{\bm \Phi}\left({\bm \Phi}^{\top}{\bm \Phi}\right)^{-1} {\bm \Phi}^{\top} \y||^2. \nonumber
\end{align}
{
{\Remark This is the unbiased estimator of the noise variance.  The classical maximum likelihood estimator in Eq. \eqref{WestML0}, i.e., $\widehat{\sigma}_e^2=\frac{1}{N} ||\y-\widehat{{\bf f}}||^2$, is instead biased.
Hence, this is an example of advantage of computing and maximizing the area under the likelihood $S({\bf y}|{\bm  \lambda})$ (i.e., the empirical Bayes approach), with respect to the classical maximum likelihood estimator.
}}
\newline
\newline
 If we replace Eq. \eqref{NoiseEst1} into \eqref{LogLikefirst},  we obtain
\begin{align}
\log S({\bf y}|{\bm \alpha}) &=-\frac{N-M}{2} -\frac{N-M}{2} \log \left(2 \pi \frac{1}{N-M} ||\y-\widehat{{\bf f}}||^2\right)-\frac{1}{2}\log\operatorname{det}\left[{\bm \Phi} ^{\top}{\bm \Phi}\right]+ const.   
\end{align}
Considering the value $M<N$ fixed and chosen by the user, we can write a cost function as
\begin{align}\label{Eq40final}
\fbox{$
C({\bm \alpha})=-\log S({\bf y}|{\bm \alpha}) = \underbrace{\frac{N-M}{2} \log\left( ||\y-\widehat{{\bf f}}||^2\right)}_{\mbox{fitting term}}+\underbrace{\frac{1}{2}\log\operatorname{det}\left[{\bm \Phi} ^{\top}{\bm \Phi}\right]}_{\mbox{penalty term}}+const,
$}
\end{align}
which is just function of ${\bm \alpha}$.
We desire to minimize the cost function $C({\bm \alpha})=-\log S({\bf y}|{\bm \alpha})$, in term of ${\bm \alpha}$.  We can clearly identify two parts:
\begin{itemize}
\item A {\it fitting term}, $\frac{N-M}{2} \log\left( ||\y-\widehat{{\bf f}}||^2\right)$, which decreases to $-\infty$ at $\y=\widehat{{\bf f}}$ (maximum overfitting). Bigger errors $||\y-\widehat{{\bf f}}||^2$ correspond to more positive values of this term. Recall that $\widehat{{\bf f}}$ depends on ${\bm \Phi}$, in fact, $\widehat{{\bf f}}={\bm \Phi}({\bm \Phi}^{\top}{\bm \Phi})^{-1} {\bm \Phi}^{\top} {\bf y}$.
\item A {\it model complexity penalization},  $\frac{1}{2}\log\operatorname{det}\left[{\bm \Phi} ^{\top}{\bm \Phi}\right]$, that penalizes the overfitting. It fosters smaller values of $\operatorname{det}\left[{\bm \Phi} ^{\top}{\bm \Phi}\right]$ (this usually happens when  ${\bm \Phi} ^{\top}{\bm \Phi}$ tends to be a full matrix), and penalizes greater values of  $\operatorname{det}\left[{\bm \Phi} ^{\top}{\bm \Phi}\right]$  (this usually happens when ${\bm \Phi} ^{\top}{\bm \Phi}$ becomes more similar to a diagonal matrix). Note that  the matrix ${\bm \Phi} ^{\top}{\bm \Phi}$ is always symmetric, and positive semi-definite \cite{Fitzgerald96_bo}. 
\end{itemize}
Minimizing Eq. \eqref{Eq40final} can be employed to tune the vector of parameters ${\bm \alpha}$   \cite{Fitzgerald96_bo,LlorenteSafePriors,LlorenteREV19}. Note also that 
\begin{align}
\exp\left(-C({\bm \alpha})\right)&=
 \frac{1}{\sqrt{\operatorname{det}\left[{\bm \Phi} ^{\top}{\bm \Phi}\right]}} \left[||\y-\widehat{{\bf f}}||^2\right]^{-\left(\frac{N-M}{2}\right)}, \nonumber Ê\\
 &=  \frac{1}{\sqrt{\operatorname{det}\left[{\bm \Phi} ^{\top}{\bm \Phi}\right]}}  \left[||\y-\widehat{{\bf f}}||\right]^{-\left(N-M\right)}, \label{Eq40final_pdf}
\end{align}
that resembles the form of a t-student density.

{\Remark  Both expressions in Eqs. \eqref{LogLikefirst}-\eqref{Eq40final} seems to be adequate for tuning the parameters of the model (Type-2 of model selection). We test them numerically in Section \ref{NumSect}.} 

{\Remark  It is possible an alternative derivation: the expressions \eqref{Eq40final}-\eqref{Eq40final_pdf} could be also obtained assuming an improper Jeffreys prior, $h(\theta)=1/\sigma_e$, and integrating out $\sigma_e$ from \eqref{aquiMLtheta} \cite[Chapter 2]{Fitzgerald96_bo}. This also confirms again that the use of improper priors is allowed in Type-2 of model selection. } 
\newline
\newline
{
The expressions \eqref{Eq40final}-\eqref{Eq40final_pdf}  can be used for tuning  ${\bm \alpha}$, i.e., we are tuning the bases $\phi_m$'s, in terms of location and scale parameters, for instance.}

\subsection{Gaussian prior over  ${\bm \theta}$}
In the previous section, we assume a improper uniform prior over  ${\bm \theta}$.
Now, let us consider a Gaussian  prior density over ${\bm \theta}$; more specifically, we assume
$$
g({\bm \theta}) =\mathcal{N}({\bm \theta}|{\bf 0}, \underbrace{{\bm \Sigma}_\theta}_{M\times M}),
$$
as a prior,  where ${\bm \Sigma}_\theta$ is a $M\times M$ covariance matrix decided and/or tuned by the user.  This prior is related to the so called {\it Tikhonov's regularization} in least squares problems. We recall that the likelihood function does not change and it is again given in Eq. \eqref{likefunctionEq}, i.e., $\ell({\bf y}| {\bm \theta})=\mathcal{N}({\bf y}| {\bm \Phi} {\bm \theta}, \sigma_e^2 {\bf I}_N)$.
\subsubsection{Posterior of ${\bm \theta}$} 

It is possible to show that the posterior density of ${\bm \theta}$ is distributed again as Gaussian density \cite{bishop2006pattern,Fitzgerald96_bo},
\begin{eqnarray}
\post({\bm \theta}|{\bf y}) =\mathcal{N}({\bm \theta}| {\bm \mu}_{\theta|y}, {\bm \Sigma}_{\theta|y}),
\end{eqnarray}
where the mean vector is 
\begin{eqnarray}
\widehat{{\bm \theta}}={\bm \mu}_{\theta|y}&=&\frac{1}{\sigma_e^2}\left(\frac{1}{\sigma_e^2}{\bm \Phi}^\top{\bm \Phi}+{\bm \Sigma}_\theta^{-1}\right)^{-1}{\bm \Phi}^\top\y,  \nonumber \\
&=&\left({\bm \Phi}^\top{\bm \Phi}+\sigma_e^2{\bm \Sigma}_\theta^{-1}\right)^{-1}{\bm \Phi}^\top\y,   \nonumber\\
&=& {\bm \Sigma}_\theta{\bm \Phi}^{\top} \left({\bm \Phi}{\bm \Sigma}_\theta{\bm \Phi}^\top+\sigma_e^2 {\bf I}_N\right)^{-1}\y,
\end{eqnarray}
 and  the covariance matrix is 
\begin{eqnarray}
{\bm \Sigma}_{\theta|y}&=&\left(\frac{1}{\sigma_e^2}{\bm \Phi}^\top{\bm \Phi}+{\bm \Sigma}_\theta^{-1}\right)^{-1}, \nonumber \\
 &=& \sigma_e^2\left({\bm \Phi}^\top{\bm \Phi}+\sigma_e^2{\bm \Sigma}_\theta^{-1}\right)^{-1}, \nonumber \\
 &=&{\bm \Sigma}_\theta-{\bm \Sigma}_\theta{\bm \Phi}^{\top}\left({\bm \Phi}{\bm \Sigma}_\theta{\bm \Phi}^{\top}+\sigma_e^2{\bf I}_N\right)^{-1}{\bm \Phi}{\bm \Sigma}_\theta,
\end{eqnarray}
where in the last expression we have used the matrix identity in \cite{bishop2006pattern}.

\subsubsection{Posterior of $f(\x)$}
Recall that the assumed model is $f(\x)={\bm \phi}({\bf x})^{\top} {\bm \theta}$ and we consider ${\bm \theta} \sim p({\bm \theta}| {\bf y})= \mathcal{N}({\bm \theta}|{\bm \mu}_{\theta|y},{\bm \Sigma}_{\theta|y})$, after seeing the data. For a fixed $\x$,  
the hidden function $f(\x)$ is again a random variable with a Gaussian posterior density, 
\begin{align}
p(f(\x) | {\bf y})= \mathcal{N}(f(\x)|\mu_{f|y}(\x),\sigma_{f|y}^2(\x)),
\end{align}
with  mean at ${\bf x}$,
\begin{align}
\mu_{f|y}(\x)=\widehat{f}(\x)&={\bm \phi}({\bf x})^{\top} \widehat{{\bm \theta}}  \nonumber \\
&={\bm \phi}({\bf x})^{\top}({\bm \Phi}^{\top}{\bm \Phi} +\sigma_e^2\boldsymbol{\Sigma}_\theta^{-1})^{-1} {\bm \Phi}^{\top} {\bf y},  \nonumber \\
&={\bm \phi}({\bf x})^{\top}\boldsymbol{\Sigma}_\theta{\bm \Phi}^{\top} \left({\bm \Phi}\boldsymbol{\Sigma}_\theta{\bm \Phi}^\top+\sigma_e^2 {\bf I}_N\right)^{-1}{\bf y},   \label{fest_basesGauss}
\end{align}
and variance
\begin{align}
\sigma_{f|y}^2(\x)&={\bm \phi}({\bf x})^{\top}  {\bm \Sigma}_{\theta|y}{\bm \phi}({\bf x}), \nonumber\\
&={\bm \phi}({\bf x})^{\top}  \left(\frac{1}{\sigma_e^2}{\bm \Phi}^\top{\bm \Phi}+\boldsymbol{\Sigma}_\theta^{-1}\right)^{-1}{\bm \phi}({\bf x}), \label{Eq41varAquiGaussPhi} \\
 &={\bm \phi}({\bf x})^{\top}{\bm \Sigma}_\theta{\bm \phi}({\bf x})- {\bm \phi}({\bf x})^{\top}{\bm \Sigma}_\theta{\bm \Phi}^{\top}\left({\bm \Phi}{\bm \Sigma}_\theta{\bm \Phi}^{\top}+\sigma_e^2{\bf I}_N\right)^{-1}{\bm \Phi}{\bm \Sigma}_\theta{\bm \phi}({\bf x}). \nonumber
\end{align}
where we have considered  ${\bm \theta}$ is distributed as its posterior, $p({\bm \theta}| {\bf y})= \mathcal{N}({\bm \theta}|{\bm \mu}_{\theta|y},{\bm \Sigma}_{\theta|y})$ and the matrix identities given in \cite{bishop2006pattern,Fitzgerald96_bo}.
The regression function is the mean solution, i.e., $\widehat{f}(\x)=\mu_{f|y}(\x)$ \cite{LucaJesseInfFus21}.
\newline
\newline
{\bf Posterior of the vector ${\bf f}$}. In the smoothing case, we have 
 ${\bf f}={\bm \Phi} {\bm \theta}$. Moreover, recall that the posterior of ${\bm \theta}$ is $ p({\bm \theta}| {\bf y})= \mathcal{N}({\bm \theta}|{\bm \mu}_{\theta|y},{\bm \Sigma}_{\theta|y})$,  hence we obtain that
\begin{eqnarray}
p({\bf f}|{\bf y}) =\mathcal{N}({\bf f}| {\bm \mu}_{f|y}, {\bm \Sigma}_{f|y}),
\end{eqnarray}
where the mean vector is 
\begin{eqnarray}
\widehat{{\bf f}}={\bm \mu}_{f|y}&=&{\bm \Phi} \widehat{{\bm \theta}}, \nonumber \\
 &=& {\bm \Phi}\left({\bm \Phi}^\top{\bm \Phi}+\sigma_e^2{\bm \Sigma}_\theta^{-1}\right)^{-1}{\bm \Phi}^\top\y, \nonumber    \\
 &=&{\bm \Phi}{\bm \Sigma}_\theta{\bm \Phi}^\top \left({\bm \Phi}{\bm \Sigma}_\theta{\bm \Phi}^\top+\sigma_e^2 {\bf I}_N\right)^{-1}\y,  \label{aquiF2stranoEq}
 \end{eqnarray}
 and the covariance matrix is
 \begin{eqnarray}
{\bm \Sigma}_{f|y}&=&{\bm \Phi}{\bm \Sigma}_{\theta|y}{\bm \Phi}^{\top},  \nonumber\\
 &=&{\bm \Phi}\left(\frac{1}{\sigma_e^2}{\bm \Phi}^\top{\bm \Phi}+{\bm \Sigma}_\theta^{-1}\right)^{-1}{\bm \Phi}^{\top}, \nonumber \\
 &=&\left[\left({\bm \Phi}{\bm \Sigma}_\theta{\bm \Phi}^{\top}  \right)^{-1}+\left(\sigma_e^2 {\bf I}_N \right)^{-1}\right]^{-1}, \nonumber   \\
 &=&{\bm \Phi}{\bm \Sigma}_\theta{\bm \Phi}^{\top}- {\bm \Phi}{\bm \Sigma}_\theta{\bm \Phi}^{\top}\left(\sigma_e^2{\bf I}_N+{\bm \Phi}{\bm \Sigma}_\theta{\bm \Phi}^{\top}\right)^{-1}{\bm \Phi}{\bm \Sigma}_\theta{\bm \Phi}^{\top},
  \end{eqnarray}
where in the last expression we have used   the matrix identity in \cite{bishop2006pattern}.


\subsubsection{Marginal likelihood}

Since  we have used a proper prior density $g({\bm \theta})$, we can compute the integral $\int_{\Theta} \ell(\y|{\bm \theta}) g({\bm \theta}) d{\bm \theta}$ without any arbitrary constant, and it can be interpreted as  a marginal likelihood  $Z=p(\y)$. Due to the assumed observation model, $\y={\bf f}+\e$, we can observe that the vector $\y$ is the sum of two independent Gaussian vectors, ${\bf f}$ and $\e$,  where
\begin{equation}
p({\bf f})=\mathcal{N}({\bf f}|{\bf 0}, {\bm \Phi}{\bm \Sigma}_\theta {\bm \Phi}^{\top}) \quad \mbox{ and }  \quad p(\e)=\mathcal{N}({\bf e}|{\bf 0}, \sigma_e^2 {\bf I}_N).
\end{equation}
The first density  $p({\bf f})$ is induced by the prior over ${\bm \theta}$, the second density $p(\e)$ is given by assumption. 
Thus, $\y$ is also distributed as a Gaussian density with mean the sums of the means, and covariance matrix the sum of the two covariance matrices, i.e.,
\begin{eqnarray}\label{MargLikeEqRegBases}
Z=p({\bf y})=p({\bf y}|{\bm  \alpha},\sigma_e,{\bm \Sigma}_\theta) = \mathcal{N}({\bf y}| {\bf 0}, {\bm \Phi}{\bm \Sigma}_\theta{\bm \Phi}^{\top}+\sigma_e^2 {\bf I}_N),
\end{eqnarray}
where we recall that a complete notation would be $p({\bf y})=p({\bf y}|{\bm \Phi},{\bm \alpha},\sigma_e,{\bm \Sigma}_\theta)$, but considering fixed the bases (hence ${\bm \Phi}$), we study the marginal likelihood as function of the hyper-parameters ${\bm \alpha}$, $\sigma_e$ and ${\bm \Sigma}_\theta$.
Therefore, the minus log-marginal likelihood $-\log Z$ is 
\begin{eqnarray}
-\log p({\bf y}|{\bm \alpha},\sigma_e,{\bm \Sigma}_\theta)&=& \nonumber\\
&=&\frac{1}{2}{\bf y}^{\top}({\bm \Phi}{\bm \Sigma}_\theta{\bm \Phi}^{\top}+\sigma_e^2 {\bf I}_N)^{-1}{\bf y}  + \frac{1}{2}\log \det\left[{\bm \Phi}{\bm \Sigma}_\theta{\bm \Phi}^{\top}+\sigma_e^2 {\bf I}_N\right]   + \frac{N}{2}\log 2\pi, \nonumber 
\end{eqnarray}
and finally 
\begin{eqnarray}
\fbox{$-\log p({\bf y}|{\bm \alpha},\sigma_e,{\bm \Sigma}_\theta)=\underbrace{\frac{1}{2}{\bf y}^{\top}({\bm \Phi}{\bm \Sigma}_\theta{\bm \Phi}^{\top}+\sigma_e^2 {\bf I}_N)^{-1}{\bf y}}_{\mbox{fitting term}}+ \underbrace{ \frac{1}{2}\log \det\left[{\bm \Phi}{\bm \Sigma}_\theta{\bm \Phi}^{\top}+\sigma_e^2 {\bf I}_N\right]}_{\mbox{penalty term}} +const., $} \label{LOG_LIKE_GAUSS}
\end{eqnarray}
The factor ${\bf y}^{\top}({\bm \Phi}{\bm \Sigma}_\theta{\bm \Phi}^{\top}+\sigma_e^2 {\bf I}_N)^{-1}{\bf y}$ is a {\it fitting term}. The second term is a {\it penalization of the model complexity}, as we have already discussed above.
Generally, one can try to maximize $p({\bf y}|{\bm \alpha},\sigma_e,{\bm \Sigma}_\theta)$, or minimize $-\log p({\bf y}|{\bm \alpha},\sigma_e,{\bm \Sigma}_\theta)$, in order to learn ${\bm \alpha}$, $\sigma_e$ and ${\bm \Sigma}_\theta$ \cite{LlorenteREV19,LlorenteSafePriors,bayesian_choice}. In this case, the covariance matrix  ${\bm \Sigma}_\theta$ is a parameter of the prior density over ${\bm \theta}$.  

\subsubsection{Trying to come back to the case of the improper uniform prior}\label{TryingBackSect}
For simplicity, we assume that the covariance matrix of the prior of ${\bm \theta}$ is diagonal, i.e., 
 ${\bm \Sigma}_\theta=\sigma_p^2 {\bf I}_N$ so that, replacing in Eq. \eqref{LOG_LIKE_GAUSS}, 
\begin{align*}
-\log Z=-\log p({\bf y}|{\bm \alpha},\sigma_e,{\bm \Sigma}_\theta)=& \\
=&\frac{1}{2}{\bf y}^{\top}\frac{1}{\sigma_p^2}\left({\bm \Phi}{\bm \Phi}^{\top}+\frac{\sigma_e^2}{\sigma_p^2} {\bf I}_N\right)^{-1}{\bf y} +\frac{1}{2}\log \det\left[\sigma_p^2{\bm \Phi}{\bm \Phi}^{\top}+\sigma_e^2 {\bf I}_N\right]+const.,
\end{align*}
First, we focus on the first term. Applying the following Woodbury matrix identity \cite{Hager89},
$$
\left(\mathbf{A}+\mathbf{C B C}^\top\right)^{-1}=\mathbf{A}^{-1}-\mathbf{A}^{-1} \mathbf{C}\left(\mathbf{B}^{-1}+\mathbf{C}^\top \mathbf{A}^{-1} \mathbf{C}\right)^{-1} \mathbf{C}^\top \mathbf{A}^{-1},
$$
to the first term, i.e.,
$$
\left(a\mathbf{I}_N+{\bm \Phi}{\bm \Phi}^{\top}\right)^{-1},
$$
where $a=\frac{\sigma_e^2}{\sigma_p^2}$, ${\bf A}=a\mathbf{I}_N$, ${\bf B}={\bf I}_N$ and ${\bf C}={\bm \Phi}$, then we obtain
\begin{align*}
\frac{1}{2\sigma_p^2}{\bf y}^{\top}\left(a\mathbf{I}_N+{\bm \Phi}{\bm \Phi}^{\top}\right)^{-1}{\bf y}&=\\
&=\frac{1}{2\sigma_p^2}{\bf y}^{\top}\left(a^{-1}{\bf I}_N-a^{-1}{\bm \Phi}\left({\bf I}_N+a^{-1}{\bm \Phi}^{\top}{\bm \Phi}\right)^{-1}{\bm \Phi}^{\top}a^{-1} \right){\bf y} \\
&=\frac{1}{2\sigma_p^2}{\bf y}^{\top}\left(a^{-1}{\bf I}_N-a^{-1}{\bm \Phi}\left(a{\bf I}_N+{\bm \Phi}^{\top}{\bm \Phi}\right)^{-1}{\bm \Phi}^{\top}\right){\bf y} \\
&=\frac{1}{2\sigma_p^2}a^{-1}\left({\bf y}^{\top}{\bf y}-{\bf y}^{\top}{\bm \Phi}\left(a{\bf I}_N+{\bm \Phi}^{\top}{\bm \Phi}\right)^{-1}{\bm \Phi}^{\top}{\bf y}\right) 
\end{align*}
Replacing $a=\frac{\sigma_e^2}{\sigma_p^2}$, we have
\begin{align}
\frac{1}{2\sigma_p^2}{\bf y}^{\top}\left(a\mathbf{I}_N+{\bm \Phi}{\bm \Phi}^{\top}\right)^{-1}{\bf y}=\frac{1}{2\sigma_e^2}\left({\bf y}^{\top}{\bf y}-{\bf y}^{\top}{\bm \Phi}\left(\frac{\sigma_e^2}{\sigma_p^2}{\bf I}_N+{\bm \Phi}^{\top}{\bm \Phi}\right)^{-1}{\bm \Phi}^{\top}{\bf y}\right).
\end{align}
Finally, for $\sigma_p^2 \rightarrow \infty$, we get
\begin{align}
 \lim_{\sigma_p^2  \rightarrow \infty} \quad \frac{1}{2\sigma_e^2}{\bf y}^{\top}\left(\frac{\sigma_e^2}{\sigma_p^2}\mathbf{I}_N+{\bm \Phi}{\bm \Phi}^{\top}\right)^{-1}{\bf y} \quad&= \frac{1}{2\sigma_e^2}\left({\bf y}^{\top}{\bf y}-{\bf y}^{\top}{\bm \Phi}\left(0 \cdot {\bf I}_N+{\bm \Phi}^{\top}{\bm \Phi}\right)^{-1}{\bm \Phi}^{\top}{\bf y}\right), \nonumber \\
&=\frac{1}{2\sigma_e^2}\left({\bf y}^{\top}{\bf y}-{\bf y}^{\top}{\bm \Phi}\left({\bm \Phi}^{\top}{\bm \Phi}\right)^{-1}{\bm \Phi}^{\top}{\bf y}\right) \nonumber \\
 &= \frac{1}{2\sigma_e^2}\left( {\bf y}^{\top}{\bf y}-  \widehat{{\bf f}}^{\top}\widehat{{\bf f}}\right), \nonumber \\
 &= \frac{1}{2\sigma_e^2} ||{\bf y}- \widehat{{\bf f}}||^2,  \label{SuperMaraVi}
\end{align}
where we have employed that $\widehat{{\bf f}}={\bm \Phi}({\bm \Phi}^{\top}{\bm \Phi})^{-1} {\bm \Phi}^{\top} {\bf y}$ as in Eq. \eqref{aquiF2stranoEq} when $\sigma_p^2  \rightarrow \infty$,  becoming equal to Eq. \eqref{Fsmoothing}, and we have used
the expression in \eqref{superIdentityEqVamos}-\eqref{superIdentityEqVamos2}.  Moreover, in the last equality, we have used Eq. \eqref{SuperRes}. Hence, as $\sigma_p^2 \rightarrow \infty$, we are able to recover the first term in  Eq. \eqref{LogLikefirst}.
\newline
\newline
Let consider now the second term as $\sigma_p^2 \rightarrow \infty$. We obtain
\begin{align}\label{AQUIsecpart}
\lim_{\sigma_p^2 \rightarrow \infty} \frac{1}{2}\log \det\left[\sigma_p^2{\bm \Phi}{\bm \Phi}^{\top}+\sigma_e^2 {\bf I}_N\right]
&=\infty,
\end{align}
so that, for the complete expression in \eqref{LOG_LIKE_GAUSS}, we have  $-\log Z \rightarrow \infty$ and hence $ Z \rightarrow 0$. 

{\Remark\label{R12} Namely, as discussed in Section \ref{vaguePriorSects},  making the Gaussian prior more diffuse penalizes  more the corresponding model and its marginal likelihood decreases to 0 (since $S<\infty$ always in this model).  }

{\Remark\label{R13} Moreover, we are not able to recover completely the improper prior case. We are just able to recover the fitting term of the $-\log S$ in Eq. \eqref{LogLikefirst} as asymptotic case of $-\log Z$ in Eq. \eqref{LOG_LIKE_GAUSS}, as shown in Eq. \eqref{SuperMaraVi}. This confirms the fact that improper priors are not allowed for computing the evidence $Z$, and that the  the area under the likelihood $S$ cannot be considered an asymptotical special case of a marginal likelihood.}

\section{Numerical examples}\label{NumEspALL}

{
The aim of this section is to clarify the use and the meaning of the generic equations and concepts described previously. We provide two examples that can help the interested reader and practitioner to apply the expressions and ideas of the work. For instance, the first example deals with a very simple scenario of estimation of the mean of a Gaussian density given $N$ data, in a Bayesian setting. However, even this simple example can serve as a guide for a proper use of the more general formulas given above, and to provide some interesting theoretical insights. The second example addresses a more general regression problem.

\subsection{First numerical example}\label{NumSect0}

In order to clarify the contents in Sections \ref{vaguePriorSects} and \ref{TryingBackSect}, we consider a simple univariate linear model with a Gaussian prior over $\theta$. This is also a very interesting special case of all the general expressions for the Bayesian regression. Indeed, le us consider  
\begin{align}
y_n=\theta+e_n,  \qquad n=1,...,N, 
\end{align}
where $e_n\sim \mathcal{N}(\epsilon|0, \sigma_e^2)$, so that the  likelihood of only one observation is
\begin{align}
p(y_n|\theta,\sigma_e)&=\frac{1}{\sqrt{2\pi \sigma_e^2}}\exp\left(-\frac{(y_n-\theta)^2}{2\sigma_e^2}\right), \quad \mbox{ and } \\
g(\theta)&= \frac{1}{\sqrt{2\pi \sigma_\theta^2}}\exp\left(-\frac{(\theta-\mu_\theta)^2}{2\sigma_\theta^2}\right),
\end{align}
is the prior  over $\theta$ with hyper-parameters $\mu_\theta$ and $\sigma_\theta^2$, hence, here ${\bm \Sigma}_\theta=\sigma_\theta^2 {\bf I}_N$.
Let us consider to observe $N$ i.i.d. data $y_i$ and let recall ${\bf y}=[y_1,....,y_N]^{\top}$.  The complete likelihood is 
\begin{align}
\ell({\bf y}|\theta,\sigma_e)=\prod_{i=1}^N p(y_n|\theta,\sigma_e).
\end{align}
In a vectorial form, we can write
\begin{align}\label{Ex0eq}
{\bf y}&={\bf 1}_N \theta+{\bf e},  
\end{align}
where ${\bf 1}_N=[1,...,1]^{\top}$ is $N\times 1$ vector of ones. If we compare Eq. \eqref{Ex0eq} with the Eq. \eqref{estoMeint2}, that we rewrite below,
$$
\underbrace{{\bf y}}_{N \times 1}=\underbrace{{\bm \Phi}}_{N\times M} \underbrace{{\bm \theta}}_{M\times 1}+ \underbrace{{\bf e}}_{N \times 1},$$
we can see that $M=1$ and ${\bm \Phi}={\bf 1}_N$, in this example.
Hence, the marginal likelihood in Eq. \eqref{MargLikeEqRegBases} becomes 
\begin{eqnarray}
Z=p({\bf y})=p({\bf y}|\sigma_p, \mu_p,\sigma_e)  = \mathcal{N}({\bf y}| {\bf 0}, \sigma_p^2{\bf 1}_N{\bf 1}_N^{\top}+\sigma_e^2 {\bf I}_N),
\end{eqnarray}
where ${\bf 1}_N{\bf 1}_N^{\top}$
is a $N\times N$ matrix of $1$'s in all the entries. This has an important statistical meaning: the observed data $y_n$'s are {\it conditional independent} given $\theta$, but they are not {\it independent}. Indeed, we have that  ${\bm \Sigma}_{yy}=\sigma_p^2{\bf 1}_N{\bf 1}_N^{\top}+\sigma_e^2 {\bf I}_N$, hence
\begin{align*}
\mbox{cov}[y_n,y_j]&= \sigma_p^2, \quad \mbox{for $n\neq j$}, \\
\mbox{cov}[y_n,y_n]&=\mbox{var}(y_n)= \sigma_p^2+\sigma_e^2.
\end{align*}
Namely, the covariance matrix of ${\bf y}$ is
\begin{align}
{\bm \Sigma}_{yy}=\sigma_p^2{\bf 1}_N{\bf 1}_N^{\top}+\sigma_e^2 {\bf I}_N=
\begin{bmatrix}
\sigma_p^2+\sigma_e^2 & \sigma_p^2 & ...  & \sigma_p^2 \\
\sigma_p^2 & \sigma_p^2+\sigma_e^2  & ...  & \sigma_p^2 \\
\vdots & \vdots & \vdots  & \vdots  \\
\sigma_p^2 & \sigma_p^2 & ...  & \sigma_p^2+\sigma_e^2  
\end{bmatrix}.
\end{align}
The log-marginal likelihood in Eq. \eqref{LOG_LIKE_GAUSS} (with the difference that here the prior has non-zero mean, i.e., $\mu_p\neq0$) can be rewritten in this scenario as 
\begin{align}\label{ZotraVezcazzo}
\log Z=\log p({\bf y})= -\frac{1}{2}\left({\bf y}-{\bm \mu}_p\right)^{\top}{\bm \Sigma}_{yy}^{-1}\left({\bf y}-{\bm \mu}_p\right) - \frac{1}{2}\log[\det {\bm \Sigma}_{yy}] - \frac{N}{2}\log( 2\pi ), 
\end{align}
where we set ${\bm \mu}_p=[\mu_p, \mu_p]^{\top}$.
Recall that actually $p({\bf y})=p({\bf y}|\sigma_p, \mu_p,\sigma_e)$. 

\subsubsection{Analysis of the area under the likelihood $S({\bf y}|\sigma_e)$}

First of all, note that in this example the only auxiliary parameter in the observation model (hence the likelihood) is $\sigma_e^2$. Clearly, there is also the main variable object of the inference, i.e., $\theta$.
In a frequentist approach, we could find 
\begin{align}
\left[\widehat{\theta}_{\texttt{ML}},\widehat{\sigma_e}_{\texttt{ML}}\right]=\arg\max_{\theta,\sigma_e} \ell({\bf y}|\theta,\sigma_e).
\end{align}
In this specific scenario, we have the analytical expressions of the two estimators in close form,
\begin{align}\label{SseEq1}
\widehat{\theta}_{\texttt{ML}}=\frac{1}{N}\sum_{n=1}^N y_n, \quad
\widehat{\sigma}_{\texttt{ML}}^2=\frac{1}{N}\sum_{n=1}^N \left(y_n-\widehat{\theta}_{\texttt{ML}}\right)^2.
\end{align}
Note that $\frac{1}{N}\sum_{n=1}^N \left(y_n-\widehat{\theta}_{\texttt{ML}}\right)^2$ is a {\it biased} estimator of the variance $\sigma_e^2$.
However, if the area under the likelihood w.r.t. $\theta$ is finite,  in a more Bayesian fashion, we could also study
\begin{align}\label{Sse}
S({\bf y}|\sigma_e)=\int_{\Theta} \ell({\bf y}|\theta,\sigma_e) d \theta. 
\end{align}
If $S({\bf y}|\sigma_e) <\infty$ for each $\sigma_e$, we can maximize this function, for instance.\footnote{{Recall that, technically, $S({\bf y}|\sigma_e)$ cannot be called marginal likelihood since has been not obtained averaging likelihood values according to a proper prior. Indeed, we can see that its value does not belong the marginal likelihood values as depicted in Figure \ref{Fig1_Ex1lab}.}}
 Note that a symmetric approach could be also employed w.r.t. $\theta$, i.e., analyzing $S({\bf y}|\theta)=\int \ell({\bf y}|\theta,\sigma_e) d \sigma_e$, if finite for each value of $\theta\in \Theta$. It is possible to show that maximizing $S({\bf y}|\sigma_e)$, we obtain
 \begin{align}\label{SseEq2}
\widehat{\sigma}_{\texttt{S}}^2=\arg\max_{\sigma_e} S({\bf y}|\sigma_e)=\frac{1}{N-1}\sum_{n=1}^N (y_n-\widehat{\theta}_{\texttt{ML}})^2.
\end{align}
that is the {\it unbiased} estimator of the variance (see also Eq. \eqref{NoiseEst1}). This shows a clear benefit in integrating out $\theta$ in \eqref{Sse}. We have checked and tested numerically this result with different vectors ${\bf y}$ of $N=2$ data (we have considered symmetric data only, for simplicity; in this case $\widehat{\theta}_{\texttt{ML}}=0$). See Table \ref{Tabla2Data}.
\begin{table}[!h]	
{
	 \caption{Numerical maximization of  $\arg\max_{\theta,\sigma_e} \ell({\bf y}|\theta,\sigma_e)$ and  $\arg\max_{\sigma_e}  S({\bf y}|\sigma_e)$. The results coincide with $\widehat{\sigma}_{\texttt{ML}}^2$ and $\widehat{\sigma}_{\texttt{S}}^2$ in Eqs. \eqref{SseEq1}-\eqref{SseEq2}.  }\label{Tabla2Data}
	 \vspace{-0.2cm}
	 \footnotesize
	\begin{center}
		\begin{tabular}{|c|c||c|c|c|c|} 
		\hline 
	{\bf Estimator}	& {\bf unbiased} & ${\bf y}=[0, 0]^{\top}$	& ${\bf y}=[-1, 1]^{\top}$ & ${\bf y}=[-2, 2]^{\top}$ &  ${\bf y}=[-3, 3]^{\top}$   \\ 
			\hline 
			\hline
			$\widehat{\sigma}_{\texttt{ML}}^2$  & \xmark& 0 & 1 & 4 &  9 \\
			 &  & & & & \\
			$\widehat{\sigma}_{\texttt{S}}^2$  &  \cmark 
			 & $10^{-6}$($\approx 0$)  & 1.988 ($\approx 2$) & 8.008 ($\approx 8$) & 17.977($\approx 18$)  \\
		 \hline
		\end{tabular}
	\end{center}
	}
\end{table} 

  {
\subsubsection{Using a proper diffuse prior: asymptotic analysis $\sigma_p \rightarrow \infty$ }

For the sake simplicity, first we keep fixed $\sigma_e^2=1$, $\mu_p=2$ and consider ${\bf y}=[-2, 2]^{\top}$.  In Figure \ref{Fig1_Ex1lab}, we can see 
$$
\log Z=\log p({\bf y}|\sigma_p,\sigma_e=1,\mu_p=2)= p({\bf y}|\sigma_p),
$$
 in Eq. \eqref{ZotraVezcazzo} as function $\sigma_p$ and the area under the likelihood $S=S({\bf y}|\sigma_e=1)$. We can observe that $\log Z \rightarrow -\infty$ hence $Z \rightarrow 0$ (hence a diffuse prior, when $S$ is finite, penalizes more and more the model)  and $Z \nrightarrow S$, i.e., $S$ cannot be recovered with asymptotical arguments.
Moreover, rewriting the negative log-marginal likelihood,
\begin{align}
-\log Z&=-\log p({\bf y}|\sigma_p)= \underbrace{\frac{1}{2}\left({\bf y}-{\bm \mu}_p\right)^{\top}{\bm \Sigma}_{yy}^{-1}\left({\bf y}-{\bm \mu}_p\right)}_{\mbox{part 1}} + \underbrace{\frac{1}{2}\log[\det {\bm \Sigma}_{yy}]}_{\mbox{part 2}}+ const,
\end{align}
we can study the behavior of the first and second part. See Figure \ref{Fig2_Ex1lab}.
As  $\sigma_p \rightarrow \infty$, we can observe numerically that the first part converges to the term $\frac{1}{2\sigma_e}||{\bf y}Ê-\widehat{{\bf f}}||^2$ where here $\widehat{{\bf f}}=\widehat{\theta}=\frac{1}{N}\sum_{n=1}^N y_n$. This result coincides with  Eq. \eqref{SuperMaraVi}. The second part diverges to $\infty$, as  $\sigma_p \rightarrow \infty$, as shown in \eqref{AQUIsecpart}. As a consequence, we have $-\log Z\rightarrow \infty$ and $Z\rightarrow 0$. Finally, in Figure \ref{Fig3_Ex1lab}, we show the  log-difference between two marginal likelihood $Z_1=p({\bf y}|\sigma_p,\sigma_e=1,\mu_p=2)$ and $Z_2=p({\bf y}|\sigma_p,\sigma_e=4,\mu_p=2)$, that  converges to a constant (horizontal asymptote) as  $\sigma_p \rightarrow \infty$. This also confirm that the Bayes factor $\frac{Z_1}{Z_2}$ still contains useful statical information even when $\sigma_p \rightarrow \infty$.

\begin{figure}[h!]
   \centering
   \centerline{
   \subfigure[\label{Fig1_Ex1lab}]{\includegraphics[width=8cm]{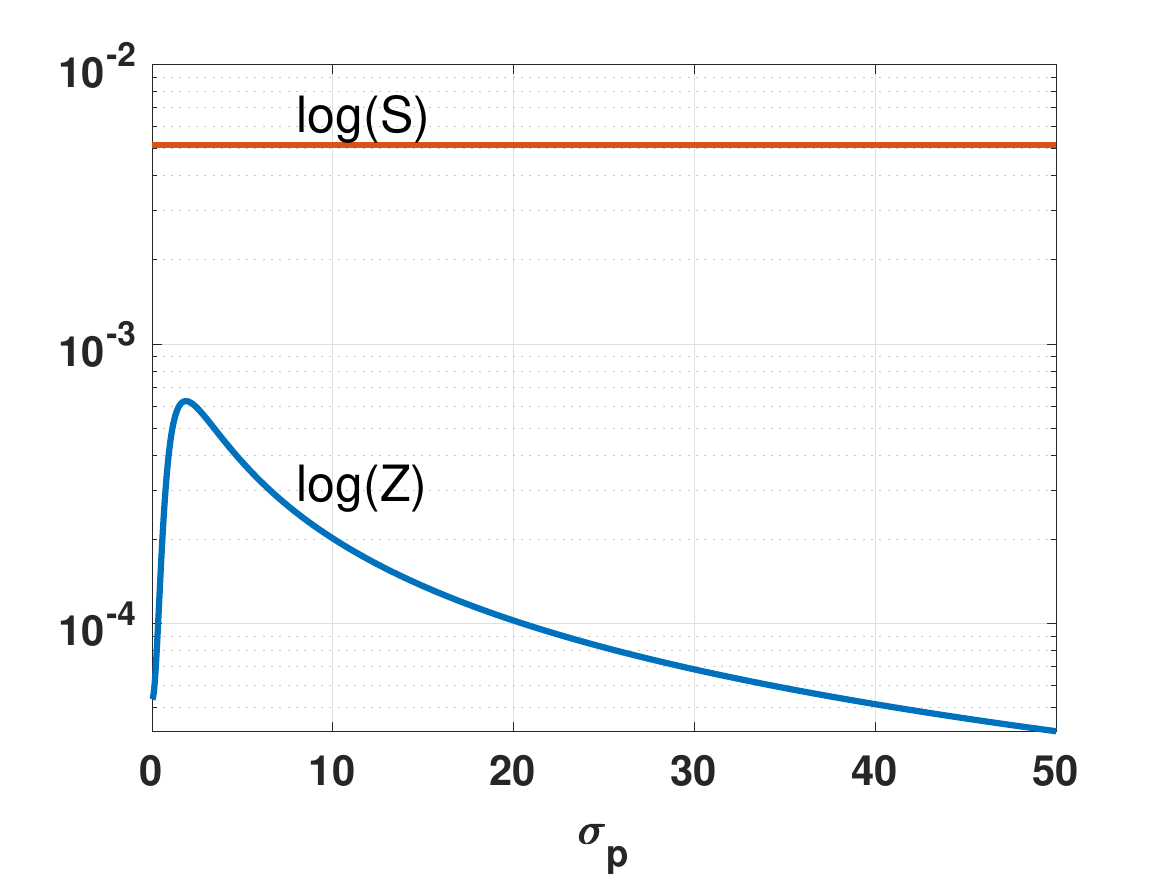} }
   \subfigure[\label{Fig2_Ex1lab}]{ \includegraphics[width=8cm]{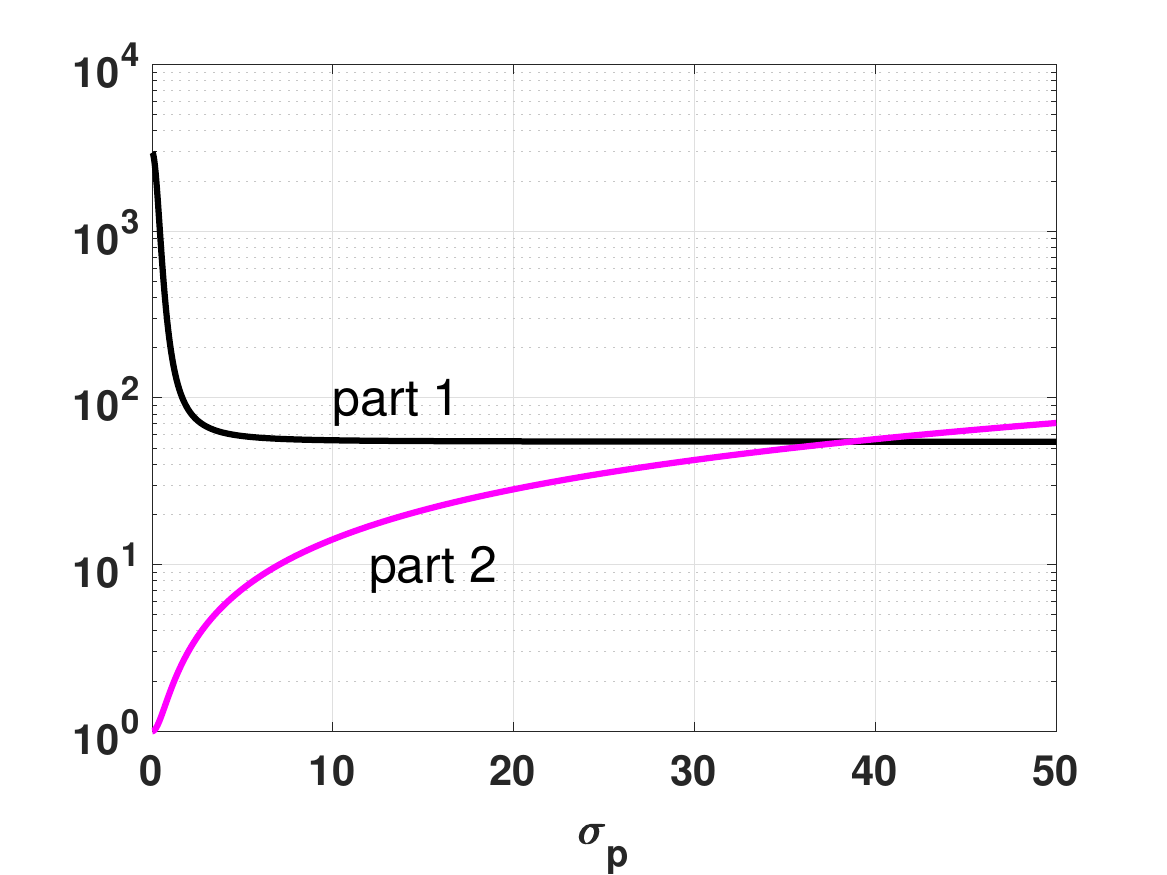} }
   }
   \centerline{
      \subfigure[\label{Fig3_Ex1lab}]{ \includegraphics[width=8cm]{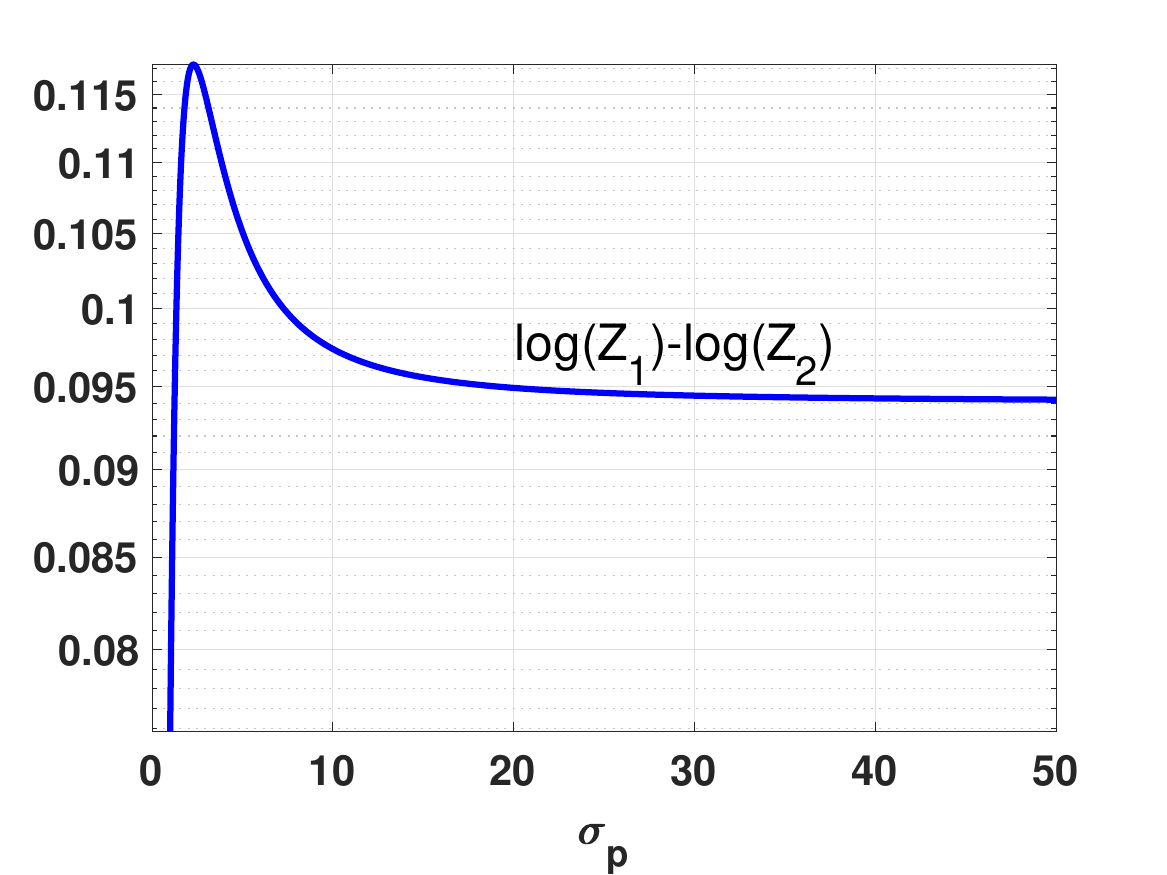} } 
      }
   \caption{{{\bf (a)} The area under the likelihood $S=S({\bf y}|\sigma_e=1)$ and  $Z$ in log-domain as function of $\sigma_p$. {\bf (b)} Part 1 in $-\log Z$ converges to a constant, whereas part 2 in $-\log Z$ diverges.    {\bf (c)} The difference  $\log Z_1 - \log Z_2$ converges to a constant (horizontal asymptote); $Z_1=p({\bf y}|\sigma_p,\sigma_e=1,\mu_p=2)$ corresponds to $\sigma_e=1$ and  $Z_2=p({\bf y}|\sigma_p,\sigma_e=4,\mu_p=2)$ corresponds to $\sigma_e=4$. }}
   \label{fig0}
\end{figure}
}
}
\subsection{{Second} numerical example}\label{NumSect}
\begin{figure}[h!]
   \centering
   \centerline{
   \subfigure[\label{dataReFig}]{\includegraphics[width=9cm]{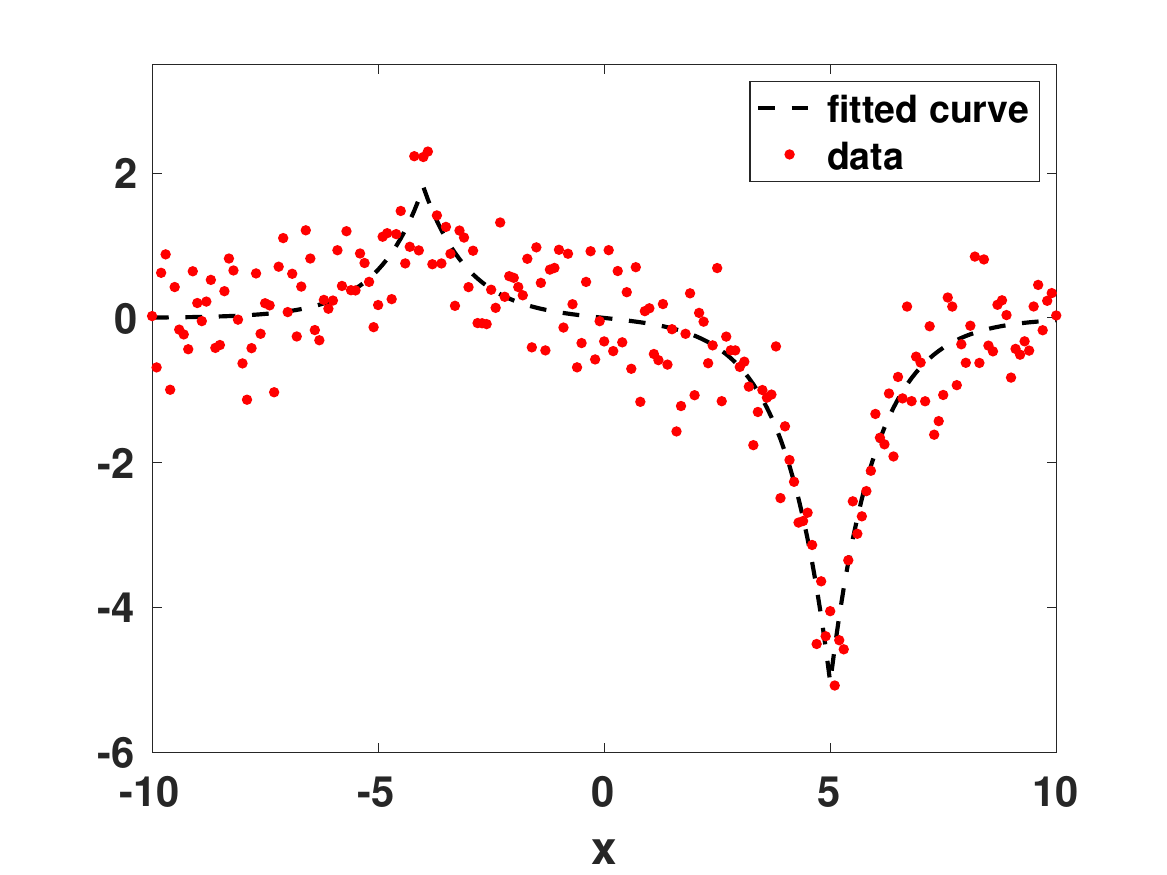} }
   \subfigure[\label{FigS}]{ \includegraphics[width=9cm]{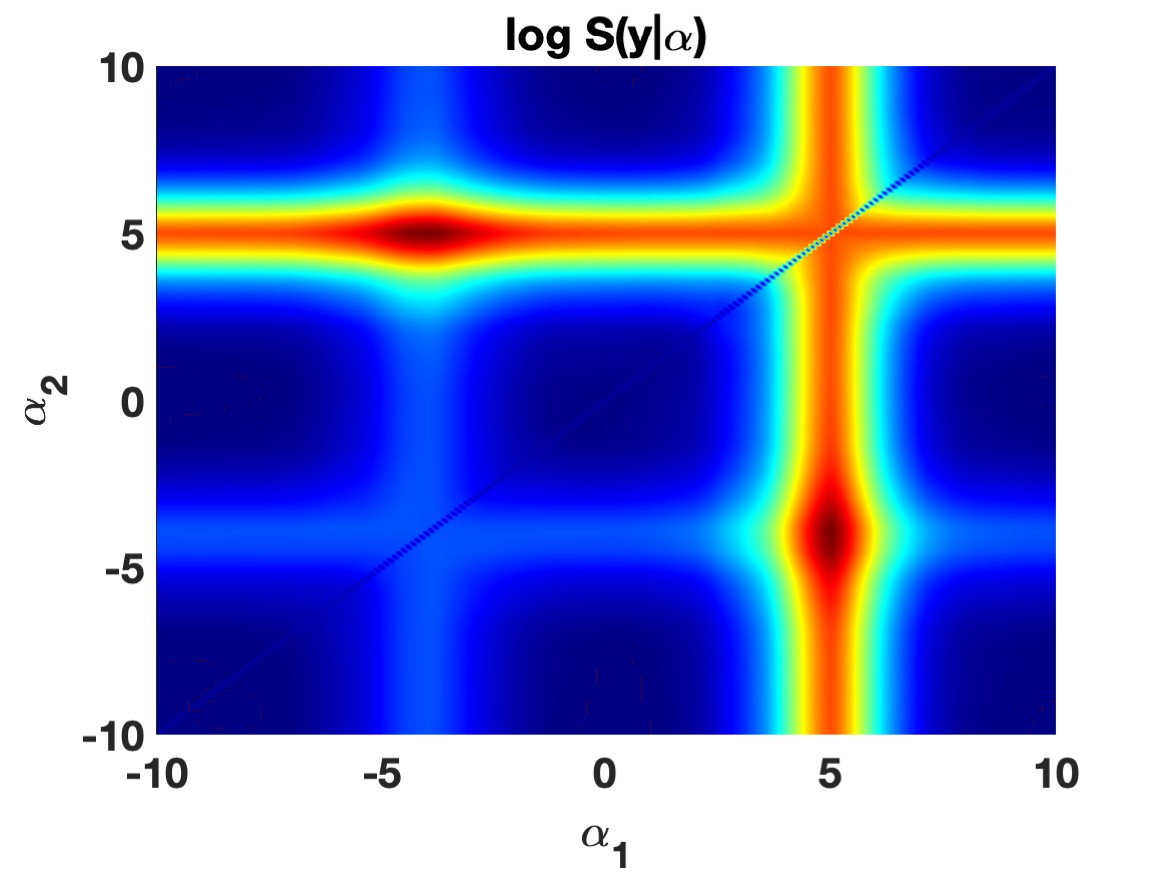} }
   }
   \caption{{\bf (a)} One realization of the data vector ${\bf y}$ and a corresponding fitted curve according to the observation model.  {\bf (b)} Example of the log area under the likelihood $\log S({\bf y}|{\bm \alpha})$ in one realization of the data vector ${\bf y}$. We can see that the maxima are localized around approximately $[-4,5]$ and $[5,-4]$ (just $[-4,5]$ is admissible since $\alpha_1<\alpha_2$). }
   \label{fig1}
\end{figure}

\begin{figure}[h!]
   \centering
      \centerline{
   \includegraphics[width=6cm]{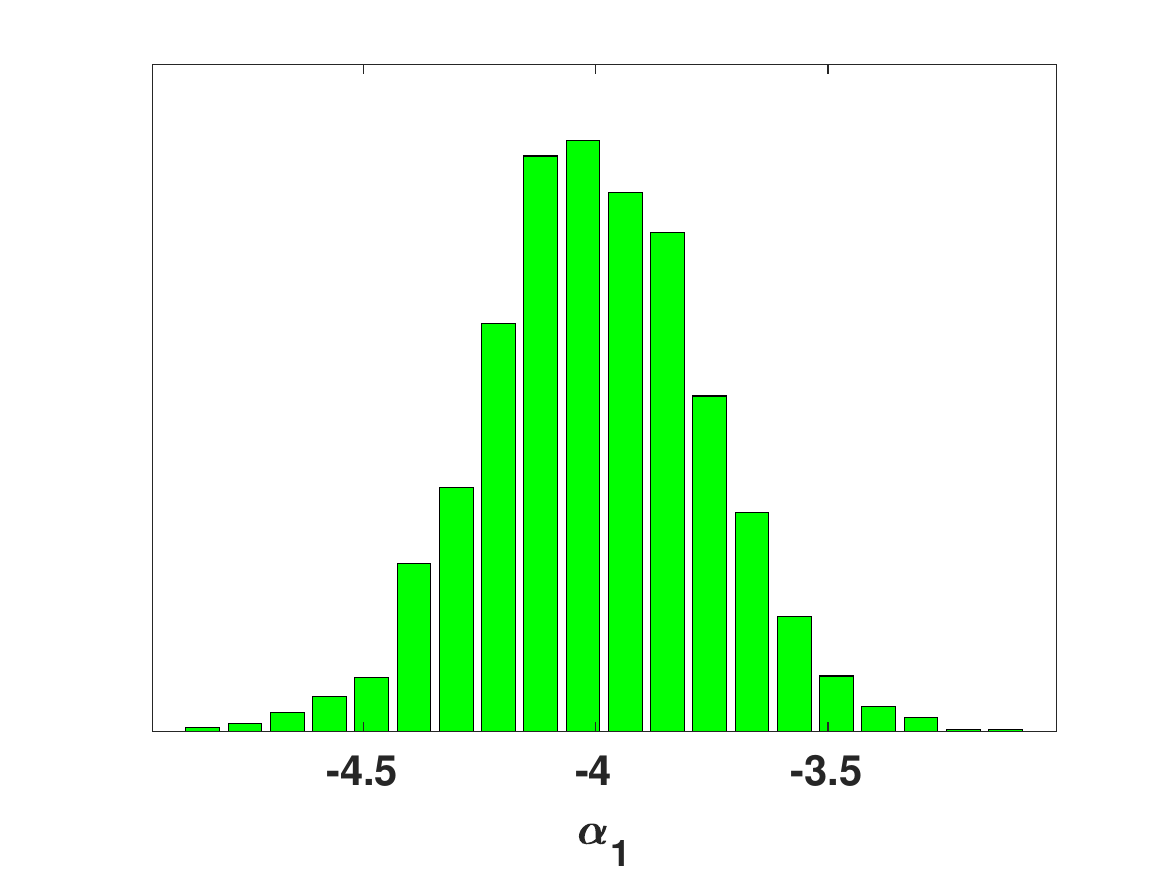} 
   \includegraphics[width=6cm]{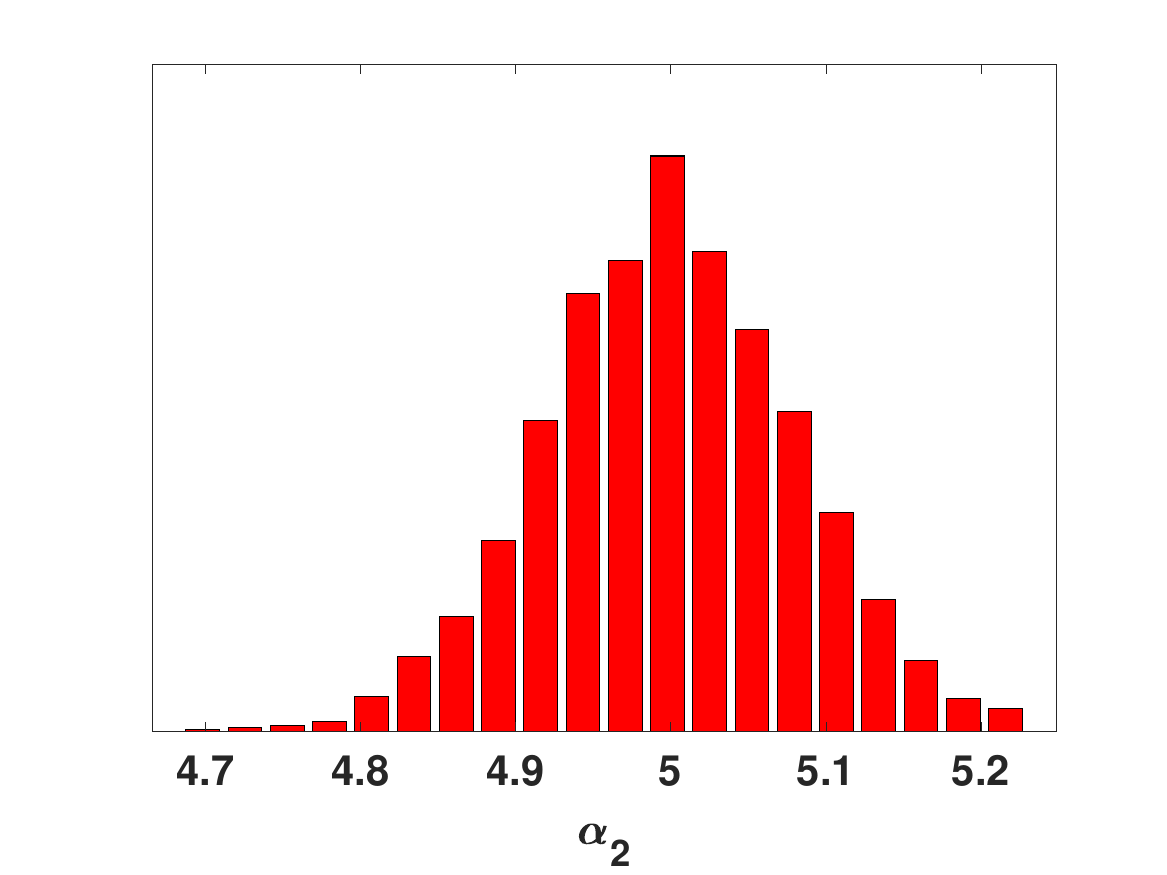} 
   \includegraphics[width=6cm]{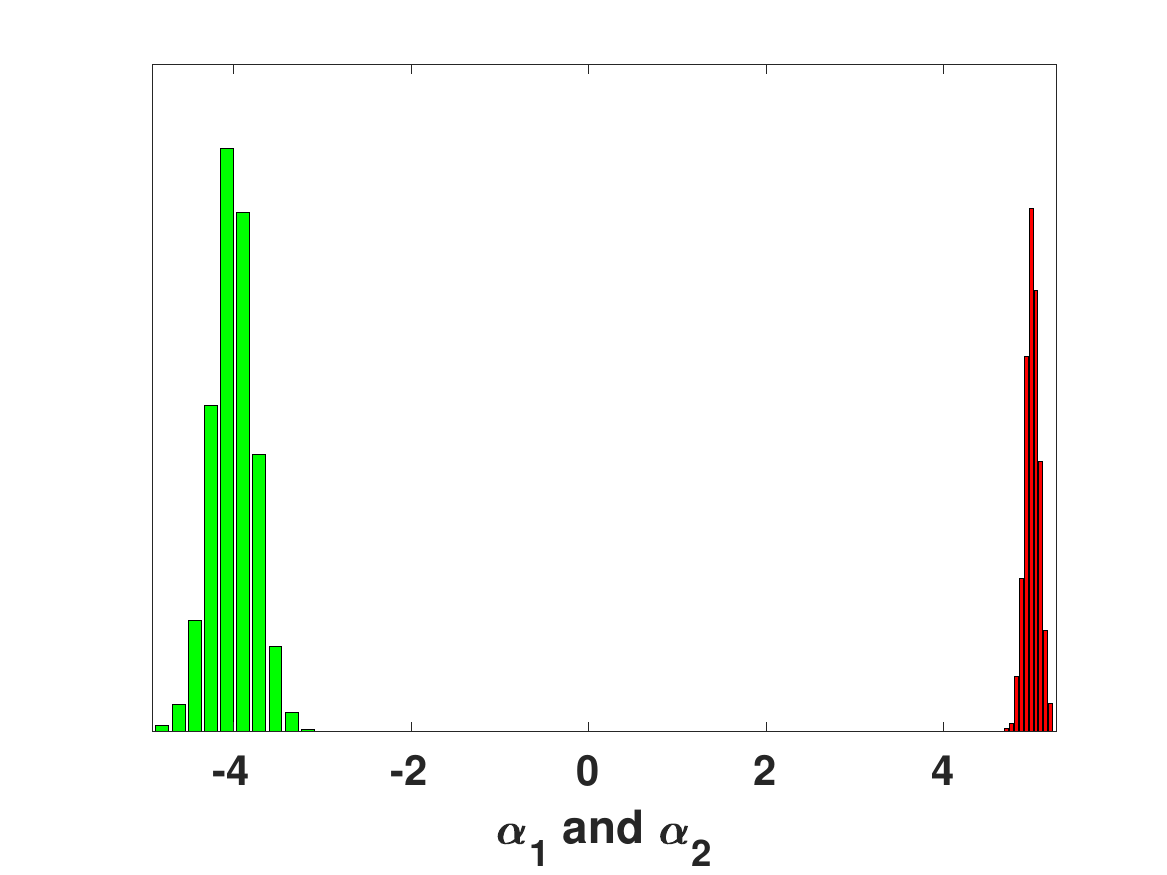} 
   }
   \caption{Histograms of  each component of estimated vector ${\bm \alpha}=[\alpha_1,\alpha_2]^{\top}$ (maximizing $S({\bf y}|{\bm \alpha})$) over $2000$ independent realizations. We can observe that bias is virtually zero and the variance in bigger for $\alpha_1$ with respect to $\alpha_2$. This is reasonable looking the realization of data in Figure \ref{dataReFig} where the (negative) pick at $x=5$ seems much more clear/evident, than the first pick at $x=-4$. }
   \label{fig2}
\end{figure}

Let us consider to observe the data points $\{x_i,y_i\}_{i=1}^N$ generated by the following model,
\begin{align} 
y_i=\theta_1\exp\left(|x_i-\alpha_1|\right)+\theta_2\exp\left(|x_i-\alpha_2|\right)+e_i, \mbox{ with } \alpha_1< \alpha_2,
\end{align} 
where $e_i\sim \mathcal{N}(e|0,\sigma_e^2)$. We set  $\sigma_e^2=0.5$, ${\bm \theta}_{\texttt{true}}=[\theta_1=2,\theta_2=-5]^{\top}$ and ${\bm \alpha}_{\texttt{true}}=[\alpha_1=-4,\alpha_2=6]^{\top}$. We consider a vector ${\bf y}=[y_1,...,y_N]^{\top}$ of $N=200$, generated by the model above considering  ${\bm \theta}_{\texttt{true}}$,  ${\bm \alpha}_{\texttt{true}}$ and equispaced values of $x_i$, from $-10$ to $10$.
\newline
\newline
 The goal is to make inference regarding ${\bm \theta}$ and ${\bm \alpha}$. We consider a Bayesian approach for inferring ${\bm \theta}$ considering an improper uniform prior. For tuning ${\bm \alpha}$, we consider the maximization of the area under the likelihood $S({\bf y}|{\bm \alpha})$ given in Section \ref{AULsect_2}. We use Eq. \eqref{West} as estimator of ${\bm \theta}$, Moreover, considering known the noise power $\sigma_e^2=0.5$, we numerically maximize $S({\bf y}|{\bm \alpha})$ in Eq. \eqref{aquiMLtheta}, or equivalently minimize \eqref{LogLikefirst} (or Eq. \eqref{Eq40final}). We compute the mean square error (MSE) in estimation for both ${\bm \theta}$ and ${\bm \alpha}$.  We repeat the procedure in $2000$ different independent runs.
\newline
\newline
The MSEs  averaged over $2000$ independent runs are $0.0271$ and $0.0317$ with respect to ${\bm \theta}$ and ${\bm \alpha}$, respectively. An example of data realization and the corresponding curve fitting is given in Figure \ref{dataReFig}. An example of log area under the likelihood $\log S({\bf y}|{\bm \alpha})$ for a given realization is shown in Figure \ref{FigS}.  The histograms of the estimated values of each component of  the vector of ${\bm \alpha}$ (in different realizations) are given in Figure \ref{fig2}.
\newline
Therefore, by this numerical example, we can confirm that the maximization of the area under the likelihood $\log S({\bf y}|{\bm \alpha})$ can be employed for tuning parameters of the observation model.

\section{Conclusions}\label{SectConcl}
In this work, we have remarked some relevant points regarding the use of diffuse priors and improper priors in level-1 and level-2 of Bayesian inference. We have stressed their impact into the inference and their possible use { with a complete discussion and by illustrative examples and  numerical experiments. The relationship with the profile likelihood approach has been also described. As a summary,  we have pointed out the following important statements:
\begin{itemize}
\item Diffuse priors are uninformative in the Level-1 of inference, but can be {\it very informative} in Level-2 (i.e., for Bayesian model selection). Indeed, if $S<\infty$, a more diffuse prior penalizes more the model providing a smaller marginal likelihood $Z$ (and asymptotically $Z\rightarrow 0$).
\item An improper prior are not allowed for the computation of the evidence $Z$.
However, we can compute the quantity $Z_{\ell \times h}$ which contains statistical information useful for  Type-2 of model selection. Namely, the  improper priors can be used when we have several models belonging to the same parametric family (i.e., for tuning parameters of a parametric model) or, more generally, for tuning parameters that are shared by different models.
\item We suggest to call the computed quantity $Z_{\ell \times h}$ as ``fake evidence'' or, in the case of uniform improper priors,  as ``the area under the likelihood'' $S=Z_{\ell \times 1}$.
\item  Another interesting aspect is that the area under the likelihood  $S$ cannot be obtained as a special asymptotic case of marginal likelihood $Z$, applying a diffuse prior and increase its scale parameter to infinity.
Namely, a diffuse prior can become asymptotically a uniform improper prior. In  Level-1 of inference, we can recover asymptotically the results obtained using a uniform improper prior, starting from a diffuse prior. In  Level-2, this is not true and $Z \nrightarrow S$, i.e., we cannot recover the area under the likelihood $S$ by a sequence of marginal likelihoods correspond to different diffuse priors (obtained increasing their scale parameter). 
\item Let assume $S<\infty$.  Even if the evidence $Z$ is not defined with an improper prior (and $Z\rightarrow 0$ using a diffuse prior), the ratio of two evidences $Z_1/Z_2$ (Bayes factor) corresponding a two different values of a shared parameter is still well-defined and can be computed by the ratio of the fake-evidences  $Z_{\ell_1\times h}/Z_{\ell_2\times h}$  (applying the same improper prior to both models).  
\end{itemize}}
\noindent
We have discussed all these aspects  firstly in a general way, and then more specifically within a Bayesian regression model, considering a uniform improper prior and a Gaussian prior. Moreover, {two numerical experiments, one involving an interesting special case and  a specific example of generalized linear model, have} been also provided performing clarifications, checks and additional remarks by numerical simulations.

\bibliographystyle{plain}
\bibliography{heretical}

\appendix


\section{Maximum likelihood estimator}\label{App1}
The maximum of the likelihood function is reached at 
\begin{equation}\label{WestML}
\widehat{ {\bm \theta}}_{\texttt{ML}}=(\underbrace{{\bm \Phi}^{\top}{\bm \Phi}}_{M\times M})^{-1} {\bm \Phi}^{\top} {\bf y}.
\end{equation}
Moreover, the covariance matrix of the estimator above is
\begin{equation}\label{Sigma_ML}
 {\bm \Sigma}_{{\bf \widehat{\theta}}}=\sigma_e^2({\bm \Phi}^{\top}{\bm \Phi})^{-1},
\end{equation}
and we can also write
\begin{equation}\label{WestMLdist}
 \widehat{ {\bm \theta}}_{\texttt{ML}}\sim \mathcal{N}(\widehat{ {\bm  \theta}}|{\bm \theta}_{\texttt{true}}, {\bm \Sigma}_{{\bf \widehat{\theta}}}),
\end{equation}
where ${\bm \theta}_{\texttt{true}}$ above represents the true vector generating the observations ${\bf y}$ following the model in Eq.\eqref{estoMeint2}, and it is also the mean of the Gaussian density above. This density is the {\it sampling distribution} of the  estimator ${\bf \widehat{\theta}}_{\texttt{ML}}$.
 Note that the sampling distribution of the estimator, $\mathcal{N}({\bf \widehat{\theta}}_{\texttt{ML}}|{\bm \theta}_{\texttt{true}}, {\bm \Sigma}_{{\bf \widehat{\theta}}})$, is philosophically completely different from a posterior distribution over the vector ${\bm \theta}$, given in Section \ref{PostTheta1}. Indeed, the sampling distribution is a probability density that describes the probabilities with which the possible values for the estimator ${\bf \widehat{\theta}}_{\texttt{ML}}={\bf \widehat{\theta}}_{\texttt{ML}}({\bf y})$ occur, when different realizations of the data ${\bf y}$ are given. 

\end{document}